\documentclass[preprint,12pt,compress]{elsarticle}

\usepackage{amsmath,amssymb,amsfonts}
\usepackage{bm}
\usepackage{graphicx}
\usepackage{xcolor}
\usepackage{empheq}
\usepackage[margin=2.4cm]{geometry}
\graphicspath{{Figures/}}
\usepackage{libertine}

\newcommand{\q}{{\bm q}}
\newcommand{\E}{{\bm E}}
\newcommand{\A}{{\bm A}}
\newcommand{\jj}{{\bm j}}
\newcommand{\br}{{\bm b}}
\newcommand{\eps}{{\bm \varepsilon}}
\newcommand{\intq}{\int_{\bm q}}
\newcommand{\tGL}{\tau_{\rm GL}}
\newcommand{\re}{\mathrm{Re}}
\newcommand{\im}{\mathrm{Im}}
\newcommand{\CE}{{\bm{\mathcal E}}}

\journal{Annals of Physics}

\begin{document}

\begin{frontmatter}

\title{Photogalvanic transport of nonreciprocal Cooper-pair fluctuations}

\author[uw,cbpf]{Joaquim Telles de Miranda}
\author[uw]{Alex Levchenko}

\address[uw]{Department of Physics, University of Wisconsin--Madison,
Madison, Wisconsin 53706, USA}
\address[cbpf]{Centro Brasileiro de Pesquisas F\'isicas (CBPF), Rio de Janeiro, Brazil}

\begin{abstract}
We develop a theory of the nonlinear optical and transport responses of
two-dimensional noncentrosymmetric superconductors in the fluctuation
regime above the transition temperature, encompassing the photogalvanic
effect, second-harmonic generation, and the photovoltaic Hall effect. In
the vicinity of the transition these responses are strongly enhanced by
preformed Cooper pairs, whose nonreciprocity enters the time-dependent
Ginzburg-Landau description in two physically distinct ways: through
thermodynamic Lifshitz invariants, which encode an asymmetric pair
spectrum, and through kinetic Lifshitz invariants, which encode an
asymmetric pair relaxation and are locked to the Langevin noise by the
fluctuation-dissipation theorem. We derive generalized master formulas
for the paraconductivity (Aslamazov-Larkin) and the quantum-interference
(Maki-Thompson) channels of the nonlinear current, valid within the
relaxational Ginzburg-Landau description of Gaussian fluctuations, for drive
frequencies below the pair-breaking scale set by temperature, and to
linear order in the nonreciprocal perturbations,
and reduce them to closed-form dimensionless functions. Circular
polarization distinguishes the mechanisms: for reciprocal
noise without momentum structure
the Aslamazov-Larkin channel is helicity independent and its
circular photogalvanic response vanishes for any pair spectrum, whereas
the Maki-Thompson channel and the nonreciprocal noise support
helicity-odd rectified currents, including a fluctuation photovoltaic
Hall current flowing transverse to the strain axis. Applications to Rashba-type ($C_{3v}$) and
Ising-type ($D_{3h}$) superconductors demonstrate how the point-group
symmetry dictates the allowed vector structures of the nonlinear
currents, and how polarization analysis together with the frequency and
dephasing dependences can be used to separate the individual channels
experimentally.
\end{abstract}


\end{frontmatter}

\tableofcontents

\section{Introduction}
\label{sec:intro}

Photoinduced transport in superconducting materials and heterostructures
has grown into a rapidly evolving frontier of modern condensed matter
physics. Three developments drive this progress. The first is the
discovery of new superconducting materials, most prominently the
two-dimensional and van der Waals superconductors in which inversion
symmetry is intrinsically broken or can be broken on demand by gating and
strain. The second is the development of experimental techniques, from intense
terahertz (THz) sources and time-resolved spectroscopies to near-field
nano-optics that access finite-momentum electrodynamics. The third is the set of
theoretical ideas rooted in the topology and quantum geometry of electronic
bands and in electronic correlations \cite{Basov2017,Torre2021,Orenstein2021}.

On the experimental side, the evidence for nonlinear optical phenomena in
superconductors is by now extensive. Terahertz second-harmonic generation
has been observed in thin films of NbN under supercurrent injection,
where the dc supercurrent supplies the inversion-symmetry breaking
\cite{Nakamura2020}. Nonlinear microwave and THz electrodynamics of
superconductors (higher-harmonic generation mediated by the Higgs
amplitude mode and by quasiparticle channels) has been established in
conventional and unconventional systems alike
\cite{Matsunaga2014,ShimanoTsuji2020,Katsumi2018}. Strongly nonlinear
responses associated with vortex motion and pinning are a classic subject
that has gained new life in the THz regime \cite{Nakamura2024,Sekiguchi2024,Hashimoto2025}. A
complementary line of experiments demonstrated various light-induced
superconducting phenomena, from the enhancement of coherence in cuprates
to metastable superconducting-like states in molecular conductors
\cite{Fausti2011,Mitrano2016,Cavalleri2018}. Superconducting fluctuations
themselves have been probed dynamically at THz frequencies
\cite{Bilbro2011}, the theory of their linear ac electrodynamics has
recently been extended to the quantum regime above the upper critical field
\cite{Zhou2026}, and the recent advent of near-field THz probes has
made it possible to visualize finite-momentum collective
electrodynamics of two-dimensional superconductors directly, as
demonstrated by the observation of a sub-gap superfluid plasmon in
few-layer Bi$_2$Sr$_2$CaCu$_2$O$_{8+x}$ \cite{vonHoegen2025}; the
theory of such near-field THz response and of the collective modes it
accesses is reviewed in Ref.~\cite{SunMillis2020}.

These experimental advances have been met by a substantial body of
theory devoted to the intrinsic nonlinear optical response of the
superconducting state itself. Ref. \cite{XMM2019}
formulated the microscopic Bogoliubov-de Gennes approach to
second-order optical effects in inversion-breaking superconductors,
demonstrating that the shift-current and second-harmonic responses are
qualitatively reorganized by the superconducting order parameter.
Ref. \cite{WatanabeDaidoYanase2022} developed the
theory of the nonreciprocal optical response of parity-breaking
superconductors at low frequencies, introducing the nonreciprocal
superfluid density and Berry-curvature-derivative contributions to the
photocurrent and second harmonic generation (SHG), and classifying the qualitative differences between
time-reversal-symmetric and time-reversal-breaking pairing states,
including the characteristic low-frequency divergences of the
condensate response. The conditions under which a nonvanishing
second-order response actually exists were made precise in Ref. \cite{TanakaWatanabeYanase2023}, who showed in a microscopic
multiband framework that the coexistence of intraband and interband
pairing components is essential; the same group extended the theory to
superconductors under magnetic fields, connecting the nonlinear response
to quantum geometry and topological superconductivity
\cite{TanakaWatanabeYanase2024}, and related multiorbital and topological
aspects were explored in Ref.~\cite{Raj2024}. Taken together, these works
reveal an uncommonly rich problem in which quasiparticle mechanisms
(shift and injection currents), condensate mechanisms (nonreciprocal
superfluid weight and its divergent low-frequency tails), collective
modes, and band-geometric and topological properties intertwine, and
they underscore that the second-order response is an exceptionally
sensitive diagnostic of broken inversion and time-reversal symmetries in
the superconducting state.

A distinct set of anomalies appears already in the normal state,
but in close proximity to the transition. Giant magnetochiral anisotropy (MCA),
 a resistance that depends on the relative orientation of current and
magnetic field, was observed in gated MoS$_2$ \cite{Wakatsuki2017} and
in SrTiO$_3$ \cite{Itahashi2020}, with the nonreciprocal signal growing
by orders of magnitude as $T\to T_c$. This behavior was traced to
fluctuating Cooper pairs: preformed pairs above $T_c$ inherit the
noncentrosymmetric band structure through odd-in-momentum terms of their
effective dispersion (Lifshitz invariants of the Ginzburg-Landau (GL)
free energy \cite{Edelstein1996,MineevSamokhin2008,Agterberg2012})
and their large susceptibility near the transition amplifies all
nonreciprocal effects. The theory of this fluctuation-enhanced
nonreciprocity was developed in
Refs.~\cite{WakatsukiNagaosa2018,Hoshino2018} and refined at the level of
nonlinear paraconductivity in Ref.~\cite{DaidoYanase2024}; the
photogalvanic response of Ising superconductors was addressed in
Ref.~\cite{Parafilo2022}, their second-harmonic generation in
Ref.~\cite{Sonowal2023}, the photovoltaic Hall effect of fluctuating
pairs in Ref.~\cite{BoevKovalev2024}, and a ratchet-type Hall
photoresponse of fluctuating pairs in spatially modulated films in
Ref.~\cite{Parafilo2025}; magnetochiral anisotropy from
fluctuations in strained transition-metal dichalcogenides and in Rashba
superconductors, as well as fluctuation-induced dichroism and gyrotropy, 
was analyzed in our companion works \cite{Levchenko2026,TdM2026a,TdM2026b,Levchenko2026TAPS}. 
This physics belongs to the broader family of
nonreciprocal transport phenomena in noncentrosymmetric conductors
\cite{Rikken2001,TokuraNagaosa2018,IdeueIwasa2021} and connects
naturally to the superconducting diode effect
\cite{Ando2020,DaidoIkedaYanase2022,Nadeem2023,Shaffer2025}.

Against this backdrop, the motivation for the present work is twofold.
First, the existing theories of the fluctuation-enhanced
photoresponse of noncentrosymmetric superconductors
\cite{Parafilo2022,Sonowal2023,BoevKovalev2024} address the
paraconductivity channel (the Aslamazov-Larkin (AL)
mechanism \cite{AslamazovLarkin1968,Schmid1966}), in which the light
couples to the center-of-mass motion of the fluctuating pairs through
the thermodynamic Lifshitz invariants of their spectrum. The
quantum-interference (anomalous Maki-Thompson, MT
\cite{Maki1968,Thompson1970}) and density-of-states contributions have
been considered for the Berry-curvature-dipole nonlinear Hall response
of time-reversal-symmetric fluctuating superconductors \cite{Dong2025}, where
they were found to be subdominant to the AL term, but not for the
photogalvanic response generated by Lifshitz invariants. In the
theory of linear fluctuation conductivity \cite{LarkinVarlamov2005} the
anomalous MT channel is of the same order as the AL term and, at weak
pair breaking, is logarithmically enhanced with respect to it. We show
that for the nonlinear response of nonreciprocal pairs the MT channel
produces photogalvanic and second-harmonic currents with the same
singularity in $T-T_c$ as the AL channel, with its own dephasing-dependent
dimensionless functions and, for circular polarization, helicity-odd
terms absent in the AL channel; the logarithmic enhancement of the linear
MT conductivity has no counterpart in the nonlinear response. Second, all prior work implicitly assumed that
nonreciprocity resides exclusively in the pair spectrum. The
Onsager reciprocity principle, however, permits the relaxation
of the fluctuating pairs to be nonreciprocal as well: the kinetic
coefficient of the time-dependent GL (TDGL) dynamics may contain
odd-in-momentum terms (kinetic Lifshitz invariants) locked
by the fluctuation-dissipation theorem (FDT) to a matching
nonreciprocity of the Langevin noise. This possibility, recently
conceptualized in Ref. \cite{Liu2026}, does not affect equilibrium
thermodynamics or linear response (we prove both statements below), but
it opens additional channels of the nonlinear response and, in
particular, makes the AL channel sensitive to the light helicity: with
FDT-locked nonreciprocal noise, circularly polarized light drives a
rectified AL current whose direction is set by the helicity. To
the best of our knowledge this mechanism was missed in all prior studies
of fluctuation photogalvanics and magnetochiral transport.

The paper is organized as follows. Section~\ref{sec:model} formulates
the extended TDGL model with thermodynamic and kinetic Lifshitz
invariants and FDT-locked noise, derives the generalized Schmid (AL)
current formula and the MT current formula, and establishes two exact
structural properties: the equilibrium state is Gibbsian and unaffected
by the kinetic invariants, and the odd-in-field part of the linear
response vanishes identically. Section~\ref{sec:ALthermo} develops the
AL channel in the thermodynamic sector: the finite-frequency expansion,
the harmonic (PGE/SHG) decomposition, the angular-reduction lemma for
cubic Lifshitz invariants, and closed-form results for the $C_{3v}$ and
$D_{3h}$ point groups. Section~\ref{sec:ALkinetic} repeats the program
for the kinetic sector of the AL channel. Section~\ref{sec:MT} treats
the MT channel in both sectors. Section~\ref{sec:PVHE} is devoted to
circular polarization and the photovoltaic Hall effect.
Section~\ref{sec:symmetry} validates all vector structures against an
exhaustive point-group analysis, and Sec.~\ref{sec:summary} summarizes
the results and outlines extensions. Technical material, the
angular-reduction proofs and our verification methodology, is
collected in the appendices.

\section{Model and main formulas}
\label{sec:model}

\subsection{TDGL dynamics with thermodynamic and kinetic Lifshitz
invariants}
\label{sec:tdgl}

We consider a two-dimensional superconductor above its transition
temperature, $T\gtrsim T_c$, described by the fluctuating pair field
$\psi_\q(t)$ with relaxational dynamics (the so-called model-A in the dissipative dynamics, see Ref. \cite{ChaikinLubensky1995}). Throughout,
$\intq\equiv\int d^2q/(2\pi)^2$ and $\hbar=k_B=1$. The Gaussian part of
the GL free energy defines the pair relaxation spectrum
\begin{equation}
\alpha_\q = Dq^2 + \Delta + \delta\alpha(\q),
\qquad
\Delta\equiv\tGL^{-1}=\frac{8(T-T_c)}{\pi},
\label{eq:alpha}
\end{equation}
where $D$ is the diffusion constant and $\delta\alpha(-\q) =
-\delta\alpha(\q)$ is the odd-in-momentum part of the spectrum (the
thermodynamic Lifshitz invariant) allowed once both inversion
and time-reversal symmetries are broken
\cite{Edelstein1996,MineevSamokhin2008,Agterberg2012}. The two point
groups of interest are:

\begin{itemize}
\item[(i)] \emph{Rashba-type superconductors} ($C_{3v}$; polar axis
$\hat z$, in-plane field $\bm B$). The leading invariant is cubic,
\begin{equation}
\delta\alpha_{C_{3v}}(\q)=\alpha_3\, q^2 (\br\cdot\q),
\qquad
\br \equiv \bm B\times\hat z,
\label{eq:LI-C3v}
\end{equation}
where $\br$ is the unique polar in-plane vector linear in $\bm B$. (A
$q$-linear invariant $\propto\br\cdot\q$ is also allowed but can be
removed by a shift of the momentum origin --- it describes helical
pairing rather than nonreciprocal transport
\cite{Agterberg2012,DaidoYanase2024}; the cubic term is the leading
non-removable invariant.)
\item[(ii)] \emph{Ising-type superconductors} ($D_{3h}$; MoS$_2$ class,
out-of-plane field $B_z$, in-plane strain $\eps$). Here
\begin{equation}
\delta\alpha_{D_{3h}}(\q)
=\kappa B_z\, q_x\big(q_x^2-3q_y^2\big)
+\eta B_z\, q^2 (\eps\cdot\q),
\qquad
\eps=(\varepsilon_{xx}-\varepsilon_{yy},\,-2\varepsilon_{xy}),
\label{eq:LI-D3h}
\end{equation}
with the first term generated by trigonal warping
[$\kappa\propto\lambda\, g\mu_B\Delta_{\rm SO}/T_c^2$ in the microscopic
band model \cite{Wakatsuki2017,TdM2026a}] and the second activated by
the $E'$ strain doublet. Both terms are $C_3$-invariant --- the strain
doublet carries angular momentum $l=-2$, so $(\eps\cdot\q)q^2\sim
e^{3i\theta}+\mathrm{c.c.}$ --- and both are odd under the vertical
mirror $x\to-x$, compensating the sign change of the pseudoscalar $B_z$.
\end{itemize}

The new element of the present theory is the observation that the
kinetic coefficient of the TDGL dynamics admits the same
classification. Let $\Gamma(\q)$ denote the (real, positive) relaxation
coefficient of the mode $\psi_\q$, with the overall relaxation constant
absorbed into the units of time so that $\alpha_\q$ is a rate. Onsager
reciprocity applied to the dissipative dynamics of a mode carrying
momentum $\q$ requires
\begin{equation}
\Gamma(\q;\bm B)=\Gamma(-\q;-\bm B),
\label{eq:onsager}
\end{equation}
so an odd-in-$\q$ part of $\Gamma$ is permitted provided it is
simultaneously odd in the time-reversal-breaking field. We write
\begin{equation}
\Gamma(\q)=1+\rho(\q),
\qquad \rho(-\q)=-\rho(\q),
\label{eq:Gamma}
\end{equation}
and call $\rho$ the kinetic Lifshitz invariant: its symmetry
classification coincides term by term with that of $\delta\alpha$. For
$C_{3v}$ the allowed drifts are
\begin{equation}
\rho_{C_{3v}}(\q)= u\,(\br\cdot\q) + u_3\, q^2(\br\cdot\q),
\label{eq:rho-C3v}
\end{equation}
and for $D_{3h}$
\begin{equation}
\rho_{D_{3h}}(\q)= \tilde\kappa B_z\, q_x\big(q_x^2-3q_y^2\big)
+\tilde\eta B_z\, q^2(\eps\cdot\q)
+\upsilon B_z\,(\eps\cdot\q).
\label{eq:rho-D3h}
\end{equation}
The linear kinetic drifts [$u$-term of
Eq.~\eqref{eq:rho-C3v} and $\upsilon$-term of Eq.~\eqref{eq:rho-D3h}]
differ from their thermodynamic counterparts in one essential respect: they
cannot be removed by the momentum-origin shift that eliminates
their thermodynamic counterparts. The shift is a property of the free
energy, it relocates the minimum of $\alpha_\q$, whereas $\rho$
parametrizes the dynamics at a given physical momentum and is not
attached to the free-energy minimum. The linear drifts are therefore
legitimate leading-order kinetic invariants, and we will see that they
produce some of the largest new effects. Microscopically, $\rho$
originates from the frequency dependence of the pair susceptibility of
the noncentrosymmetric band structure: expanding the inverse fluctuation
propagator as $L^{-1}(\q,\omega)\simeq
-\Gamma(\q)\,i\omega+\alpha(\q)+\dots$, the same
SOC$\times$field physics that generates $\delta\alpha$ in the static
part generates $\rho$ in the dynamical part, with no additional
smallness; for the Rashba case one finds $u\,\xi^{-1}\sim
\sqrt{\epsilon}$ at the fluctuation momentum scale, parametrically
comparable to the thermodynamic channel \cite{Liu2026}.
The two sectors can be compared directly because $u$ and $\alpha_3/D$
are both lengths: the dimensionless strength of the kinetic drift at the
fluctuation momentum $q\sim\xi_{\rm GL}^{-1}=\sqrt{\Delta/D}$ is
$u/\xi_{\rm GL}$, while that of the thermodynamic cubic invariant,
$\delta\alpha/\alpha\sim\alpha_3q^3/\Delta$, is $(\alpha_3/D)/\xi_{\rm GL}$.
The linear kinetic drift therefore competes with the thermodynamic
invariant when $u\sim\alpha_3/D$, and the results of
Secs.~\ref{sec:ALkinetic} and \ref{sec:MT} show that the two then
contribute to the nonlinear current at the same order in $\tGL$ with
numerical weights of order unity. In a microscopic model both lengths are
set by the same product of spin-orbit coupling and time-reversal-breaking
field, the static part of the odd pair susceptibility fixing $\alpha_3$
and its leading frequency dependence fixing $u$, so that $u\sim\alpha_3/D$
up to a model-dependent factor of order one \cite{Liu2026}. We treat
$u$, $u_3$, $\tilde\kappa$, $\tilde\eta$, $\upsilon$ as phenomenological
parameters in what follows; their quantitative values for specific
Rashba and Ising materials, including their disorder dependence, require
the microscopic calculation outlined in Sec.~\ref{sec:summary}.

The stochastic TDGL equation in the presence of the electromagnetic
potential $\A(t)$ [drive $\E(t)=-\partial_t\A$; pair charge $2e$] reads
\begin{equation}
\Gamma\big(\q-2e\A(t)\big)\,\partial_t\psi_{\q}
=-\,\alpha\big(\q-2e\A(t)\big)\,\psi_{\q}+\zeta_{\q}(t),
\label{eq:tdgl}
\end{equation}
and the classical FDT ties the Langevin noise to the same
momentum-dependent kinetic coefficient,
\begin{equation}
\big\langle\zeta_{\q}(t)\,\zeta^{*}_{\q'}(t')\big\rangle
=2T\,\Gamma\big(\q-2e\A(t)\big)\,
(2\pi)^2\delta^{(2)}(\q-\q')\,\delta(t-t'),
\label{eq:noise}
\end{equation}
the lock being applied at the gauge-invariant instantaneous momentum
(the standard adiabatic assumption of TDGL
\cite{LarkinVarlamov2005,GE1968}). Equations
\eqref{eq:tdgl}--\eqref{eq:noise} define the model completely. We work
to linear order in the odd perturbations $\{\delta\alpha,\rho\}$ and to
second order in the drive field. The analysis reveals two structural properties that shape all the physics discussed below.

\paragraph{(i) Equilibrium is untouched} At $\E=0$ the stationary
solution of Eqs.~\eqref{eq:tdgl}--\eqref{eq:noise} is
\begin{equation}
\big\langle|\psi_{\q}|^2\big\rangle_{\rm eq}=\frac{T}{\alpha_\q},
\label{eq:gibbs}
\end{equation}
pointwise independent of $\Gamma$: a mode with $\rho(\q)>0$ relaxes faster
and is pumped harder, and the two effects cancel identically.
The static distribution is Gibbsian, no equilibrium current can arise,
and no thermodynamic quantity is modified by the kinetic invariants, 
this is enforced by Onsager reciprocity and the FDT. Conversely, any
treatment that modifies the relaxation rate but keeps white noise (or
vice versa) violates the FDT and produces spurious equilibrium currents;
the friction--noise pair $(\Gamma,\,2T\Gamma)$ must be locked
throughout.

\paragraph{(ii) No linear-response signature} Expanding the response to
first order in $\E$, the odd-in-$\{\delta\alpha,\rho\}$ part of the
linear conductivity vanishes identically, in both the AL and MT channels. 
Nonreciprocity of either type is thus invisible in
equilibrium thermodynamics and in linear transport; its first observable
consequences are the second-order responses studied in this paper. The
mechanism by which the kinetic invariant eventually contributes is the
nonequilibrium mismatch of the two locked factors: along the
field-accelerated trajectory the relaxation rate and the noise are
sampled at different momenta, and this mismatch rectifies.

\subsection{Generalized Aslamazov-Larkin-Schmid formula}
\label{sec:genAL}

The AL contribution to the current is carried by the supercurrent of the
fluctuating pairs,
\begin{equation}
\jj_{\rm AL}(t)=2e\intq \partial_\q\alpha\big(\q-2e\A(t)\big)\,
\big\langle|\psi_\q(t)|^2\big\rangle,
\label{eq:jAL-def}
\end{equation}
where the vertex $2e\,\partial_\q\alpha$ follows from the free energy, 
the kinetic coefficient does not enter the current operator. The
driven second moment follows by solving the linear Langevin equation
\eqref{eq:tdgl} along the accelerated trajectory,
\begin{equation}
\psi_\q(t)=\int_{-\infty}^{t}\!\frac{dt'\;\zeta_\q(t')}
{\Gamma\big(\q-2e\A(t')\big)}\,
\exp\Big[-\!\int_{t'}^{t}\!ds\,
\frac{\alpha}{\Gamma}\big(\q-2e\A(s)\big)\Big],
\end{equation}
and averaging over the noise \eqref{eq:noise}:
\begin{equation}
\big\langle|\psi_{\q}(t)|^2\big\rangle
=2T\int^{t}_{-\infty}\!dt'\,
\Gamma^{-1}\big(\q-2e\A(t')\big)\,
\exp\Big[-2\!\int^{t}_{t'}\!ds\,
\frac{\alpha}{\Gamma}\big(\q-2e\A(s)\big)\Big].
\label{eq:psi2}
\end{equation}
Note the two distinct appearances of $\Gamma$: one factor of
$\Gamma^{-1}$ from the noise vertex [product of the noise correlator
$2T\Gamma$ and two propagator factors $\Gamma^{-1}$], and the
effective rate $\bar\alpha\equiv\alpha/\Gamma$ in the exponent.
To linear order in the odd perturbations,
\begin{equation}
\bar\alpha=\alpha_0+\delta\alpha-\alpha_0\rho,
\qquad
\Gamma^{-1}=1-\rho,
\qquad \alpha_0\equiv Dq^2+\Delta .
\label{eq:baralpha}
\end{equation}
Combining Eqs.~\eqref{eq:jAL-def} and \eqref{eq:psi2} yields the
generalized Schmid formula
\begin{empheq}{align}
\jj_{\rm AL}(t)=4eT\intq
\partial_{\q}\alpha\big(\q-2e\A(t)\big)
\int^{t}_{-\infty}\!dt'\,&
\big[1-\rho\big(\q-2e\A(t')\big)\big]
\nonumber\\[2pt]
&\times
\exp\Big[-2\!\int^{t}_{t'}\!ds\;
\bar\alpha\big(\q-2e\A(s)\big)\Big].
\label{eq:genAL}
\end{empheq}
At $\rho=0$ this is the Schmid formula
\cite{Schmid1966,DaidoYanase2024} employed in the fluctuation
literature; it is convenient to shift the time argument and use the
equivalent form
\begin{equation}
\jj_{\rm AL}=4eT\intq \partial_\q\alpha_\q\int_{-\infty}^{0}\!dt_2\,
\exp\Big[-2\!\int_{t_2}^{0}\!dt_1\,
\alpha\big(\q-2e\A(t_1{+}t)+2e\A(t)\big)\Big]
\label{eq:AL-master}
\end{equation}
in the thermodynamic sector ($\rho=0$), where the gauge-invariant
combination $\q-2e\A(t_1{+}t)+2e\A(t)$ makes the physical-momentum
content explicit. The equilibrium check~(i) is manifest in
Eq.~\eqref{eq:genAL}: at $\A=0$ the noise factor $[1-\rho]$ and the rate
shift $-\alpha_0\rho$ in the exponent cancel pointwise upon performing
the $t'$ integral, $[1-\rho]/[2(\alpha_0-\alpha_0\rho+\delta\alpha)]
=1/2\alpha_\q+\mathcal O(\rho^2)$, recovering Eq.~\eqref{eq:gibbs}.

\subsection{Maki-Thompson current formula}
\label{sec:genMT}

The quantum-interference (anomalous MT) contribution does not follow
from the TDGL equation; it must be imported from the microscopic theory.
In the Keldysh nonlinear sigma-model derivation of fluctuation transport
\cite{Liu2026,LevchenkoKamenev2007,Kamenev2011}, the anomalous MT current is
built from two electronic Cooperons attached to the Keldysh component of
the pair propagator, physically, the driven pair correlator
$\langle|\psi_\q|^2\rangle$ sandwiched between retarded and advanced
Cooperon ladders. Two structural facts fix its generalization to the
nonreciprocal problem. The Cooperons are electronic objects, controlled by
the dephasing rate $\tau_\phi^{-1}$ and independent of the order-parameter
kinetic coefficient. The order-parameter noise enters only through the
Langevin factor, that is, through the same combination of noise vertex and
dressed exponent as in Eq.~\eqref{eq:psi2}. The master formula therefore reads
\begin{empheq}{align}
\jj_{\rm MT}=16e^2TD\intq\int_{-\infty}^{t}\!dt_1\,&
\exp\Big[-2\!\int_{t_1}^{t}\!dt'\,
\beta\big(\q-e\A(t')-e\A(2t_1{-}t')\big)\Big]\,\E(2t_1{-}t)
\nonumber\\[2pt]
\times\int_{-\infty}^{t_1}\!dt_2\,&
\big[1-\rho\big(\q-2e\A(t_2)\big)\big]
\exp\Big[-2\!\int_{t_2}^{t_1}\!dt''\,
\bar\alpha\big(\q-2e\A(t'')\big)\Big],
\label{eq:genMT}
\end{empheq}
where
\begin{equation}
\beta_\q=Dq^2+\tau_\phi^{-1}
\label{eq:beta}
\end{equation}
is the Cooperon (particle-particle) pole with the pair-breaking rate
$\tau_\phi^{-1}$, and the electric field enters in three places:
through the explicit vertex $\E(2t_1{-}t)$, through the accelerated
momenta in the Langevin factor, and through the accelerated momentum
$\q-e\A(t')-e\A(2t_1-t')$ of the Cooperon pole $\beta$ in the first
exponent. The last dependence is of the same order as the second,
$\E\cdot\nabla\beta_\q=\E\cdot\nabla\alpha_\q=2eD\,\E\cdot\q$, and is
retained to first order in the field throughout; it contributes to every
finite-frequency MT kernel below and is responsible for the cancellation
of the secular terms discussed in Sec.~\ref{sec:MTsecular}. At
$\rho=\delta\alpha=0$,
Eq.~\eqref{eq:genMT} reproduces the classical anomalous MT
paraconductivity \cite{Thompson1970,LarkinVarlamov2005}; the
form of the time arguments follows from the microscopic derivation of
Ref.~\cite{LevchenkoKamenev2007} and encodes the retarded-advanced
structure of the two Cooperons. Field corrections to the electronic
current vertices and to the internal structure of the Cooperons are not
enhanced by $\tGL$ and are omitted, consistently with the Ginzburg-Landau
truncation that underlies both Eqs.~\eqref{eq:genAL} and \eqref{eq:genMT}.
The low-frequency limits of all MT results below (the static split, the
dc totals, and the cancellation of the $1/\omega$ poles) do not depend on
the time structure of the vertex and the Cooperon argument, whereas the
finite-frequency profiles do; the latter therefore rest on the
retarded-advanced structure imported from the microscopic theory.

A comment on the Cooperon pole \eqref{eq:beta} is in order.
Microscopically, the Cooperon is built from the same single-particle
Green's functions that, resummed in the particle-particle channel, form
the pair propagator. It therefore inherits from the band structure and
the symmetry-breaking fields odd-in-momentum terms $\delta\beta(\q)$ of
the same symmetry as $\delta\alpha(\q)$; see the derivation of the
disorder-dressed pair propagator in Ref.~\cite{TdM2026a}. In the present
phenomenological description the nonreciprocal parts of $\alpha_\q$ and
$\beta_\q$ are independent inputs, and the results below are given for
$\delta\beta=0$. A finite $\delta\beta$ is accommodated without new
integrals. At $\omega\to0$ the nonreciprocal MT current is linear in the
pair $(\delta\alpha,\delta\beta)$ and vanishes identically for
$\delta\beta=\delta\alpha$, since $\beta_\q-\alpha_\q$ is then a constant
and the integrand of the dc current is a total momentum gradient. The
dc results of the thermodynamic sector in Sec.~\ref{sec:MT} thus hold
with $\delta\alpha\to\delta\alpha-\delta\beta$: the MT channel measures
the mismatch between the nonreciprocity of the pair spectrum and that
of the Cooperon pole, whereas the AL channel and the kinetic sector are
not affected.

The microscopic content of the phenomenological parameter $\tau_\phi$
depends on the dominant pair-breaking mechanism (spin-flip scattering, the
orbital effect of the magnetic field, or inelastic processes); near $T_c$
it enters all MT results through the single dimensionless ratio
\begin{equation}
x\equiv \tGL/\tau_\phi,
\label{eq:x}
\end{equation}
together with the dimensionless drive frequency
\begin{equation}
\nu\equiv\omega\tGL .
\label{eq:nu}
\end{equation}
Throughout the paper the drive is monochromatic and linearly polarized,
$\E(t)=\E\cos\omega t$, in the gauge
\begin{equation}
\A(t)=-\frac{\E}{\omega}\,\sin\omega t ,
\label{eq:gauge}
\end{equation}
except in Sec.~\ref{sec:PVHE}, where elliptic polarization is treated
with two amplitudes. The observables are the harmonics of the induced
current: the rectified (dc) component, the photogalvanic effect
(PGE); and the $2\omega$ component, second-harmonic generation
(SHG), with in-phase ($\cos2\omega t$) and quadrature
($\sin2\omega t$) amplitudes.

The range of validity of the theory is set by the relaxational TDGL
description itself. The pair dynamics is adiabatic with respect to the
quasiparticles when $\omega\ll T$ (in units $\hbar=k_B=1$), which is the
condition under which the frequency dependence of the fluctuation
propagator is captured by the single relaxational term of
Eq.~\eqref{eq:tdgl}; for $T_c\sim10\,$K this restricts the drive to
frequencies below a few hundred GHz, and for higher-$T_c$ systems to the
low-THz range. At higher frequencies the quasiparticle sector and the
nonadiabatic pair dynamics studied microscopically for the
superconducting state \cite{Zachmann2013} must be included; these are
beyond the TDGL. The fluctuations are treated in the Gaussian regime,
$Gi\ll\epsilon=(T-T_c)/T_c\ll1$ with $Gi$ the Ginzburg-Levanyuk number,
so that the interaction of fluctuations and the critical region are
excluded, and the nonreciprocal couplings $\{\alpha_3,\kappa,\eta\}$ and
$\{u,u_3,\tilde\kappa,\tilde\eta,\upsilon\}$ enter as phenomenological
constants to be fixed by a microscopic theory. Finally, the expansion
to second order in the drive requires that the momentum acquired by a
pair during its lifetime be small compared with the inverse correlation
length, $eE\,\xi_{\rm GL}\tGL\ll1$ for $\omega\tGL\lesssim1$, a condition
that relaxes to $eE\sqrt{D/\omega^{3}}\ll1$ for $\omega\tGL\gg1$. The
corresponding field scale vanishes as $(T-T_c)^{3/2}$ at low frequency;
above it the response is nonperturbative in the field. All statements below are
formulated within this framework; the material-specific magnitudes discussed
in Sec.~\ref{sec:summary} require microscopic input.

\section{Aslamazov-Larkin channel: thermodynamic sector}
\label{sec:ALthermo}

In this section we set $\rho=0$ and develop the AL response generated by
the thermodynamic Lifshitz invariants, at arbitrary drive frequency. The
calculation proceeds in three steps: the exact time integrals of the
main formula \eqref{eq:AL-master}, the harmonic decomposition of the
resulting kernels, and the angular-plus-radial reduction of the momentum
integrals.

\subsection{Time integrals and harmonic decomposition}
\label{sec:ALtime}

With the gauge \eqref{eq:gauge} the momentum shift entering
Eq.~\eqref{eq:AL-master} is
\begin{equation}
\bm s(t_1)= -2e\A(t_1{+}t)+2e\A(t)
=\frac{2e}{\omega}\,\E\,\big[\sin\omega(t_1{+}t)-\sin\omega t\big],
\label{eq:shift}
\end{equation}
and we expand $\alpha(\q+\bm s)=\alpha_\q + s_i\partial_i\alpha_\q
+\tfrac12 s_is_j\partial_i\partial_j\alpha_\q+\dots$ to second order in
the field. The inner time integral is elementary but must be carried out
carefully; one finds
\begin{align}
\int_{t_2}^{0}\!dt_1\,\alpha_{\q+\bm s(t_1)}
&=-\alpha_\q t_2
+\frac{2e}{\omega^2}\Big[\cos\omega(t{+}t_2)-\cos\omega t
+\omega\, t_2\,\sin\omega t\Big]E_i\,\partial_i\alpha_\q
\nonumber\\
&\quad+\frac{e^2}{2\omega^3}\Big[\sin2\omega(t{+}t_2)+3\sin2\omega t
-8\sin\omega t\cos\omega(t{+}t_2)
\nonumber\\
&\hspace{2.2cm}
+2\omega\,t_2\cos2\omega t-4\omega\,t_2\Big]
E_iE_j\,\partial_i\partial_j\alpha_\q .
\label{eq:ALinner}
\end{align}
Exponentiating and performing the $t_2$ integral against
$e^{2\alpha_\q t_2}$ gives, to second order in $\E$,
\begin{align}
\int_{-\infty}^{0}\!\!dt_2\,
e^{-2\int_{t_2}^0 dt_1\, \alpha_{\q+\bm s(t_1)}}
&=\frac{1}{2\alpha}
+\frac{2\alpha\cos\omega t+\omega\sin\omega t}
{\alpha^2(4\alpha^2+\omega^2)}\,
e E_i\partial_i\alpha
\nonumber\\
&\hspace{-2.6cm}
+\frac{(12\alpha^2{+}\omega^2)(\alpha^2{+}\omega^2)
+(12\alpha^4{-}7\alpha^2\omega^2{-}\omega^4)\cos2\omega t}
{\alpha^3(\alpha^2+\omega^2)(4\alpha^2+\omega^2)^2}\,
e^2E_iE_j\,\partial_i\alpha\,\partial_j\alpha
\nonumber\\
&\hspace{-2.6cm}
+\frac{2\alpha\omega(10\alpha^2{+}\omega^2)\sin2\omega t}
{\alpha^3(\alpha^2+\omega^2)(4\alpha^2+\omega^2)^2}\,
e^2E_iE_j\,\partial_i\alpha\,\partial_j\alpha
\nonumber\\
&\hspace{-2.6cm}
-\frac{2\alpha^2{+}2\omega^2+(2\alpha^2{-}\omega^2)\cos2\omega t
+3\alpha\omega\sin2\omega t}
{2\alpha^2(\alpha^2+\omega^2)(4\alpha^2+\omega^2)}\,
e^2E_iE_j\,\partial_i\partial_j\alpha
+\mathcal O(E^3),
\label{eq:ALt2}
\end{align}
where $\alpha\equiv\alpha_\q$ and we used the factorization
$12\alpha^4+13\alpha^2\omega^2+\omega^4
=(12\alpha^2+\omega^2)(\alpha^2+\omega^2)$. Every term of
Eq.~\eqref{eq:ALt2} was verified symbolically and numerically (see
\ref{app:verification}). Attaching the current vertex
$4eT\,\partial_\q\alpha(\q-2e\A(t))$, whose own field dependence
contributes at the same order and is included in the harmonic
decomposition, and projecting onto the dc and $2\omega$ harmonics
yields the two finite-frequency kernels of the AL channel:
\begin{align}
\delta\jj^{\rm PGE}_{\rm AL}
&=4e^3T\,E_iE_j\intq\Big[
\underbrace{\frac{12\alpha^2+\omega^2}
{\alpha^3(4\alpha^2+\omega^2)^2}}_{\textstyle C_A^{\rm dc}}\,
\partial_i\alpha\,\partial_j\alpha
\;\underbrace{-\frac{1}{\alpha^2(4\alpha^2+\omega^2)}}_{\textstyle
C_B^{\rm dc}}\,\partial_i\partial_j\alpha
\Big]\,\partial_{\q}\alpha ,
\label{eq:ALPGE}\\[4pt]
\delta\jj^{\rm SHG}_{\rm AL}
&=4e^3T\,E_iE_j\intq\Big[
\frac{(12\alpha^4{-}7\alpha^2\omega^2{-}\omega^4)\cos2\omega t
+2\alpha\omega(10\alpha^2{+}\omega^2)\sin2\omega t}
{\alpha^3(\alpha^2+\omega^2)(4\alpha^2+\omega^2)^2}\,
\partial_i\alpha\,\partial_j\alpha
\nonumber\\&\hspace{2.9cm}
-\frac{(2\alpha^2{-}\omega^2)\cos2\omega t+3\alpha\omega\sin2\omega t}
{2\alpha^2(\alpha^2+\omega^2)(4\alpha^2+\omega^2)}\,
\partial_i\partial_j\alpha\Big]\,\partial_{\q}\alpha .
\label{eq:ALSHG}
\end{align}
Equations \eqref{eq:ALPGE}--\eqref{eq:ALSHG} are even in
$\delta\alpha\to-\delta\alpha$ at zeroth order and odd at first order:
the photogalvanic and second-harmonic currents are linear in the
Lifshitz invariant, as dictated by symmetry. For reference, the linear
response extracted from the same expansion reproduces the classical
paraconductivity result: at $\omega\to0$,
\begin{equation}
\jj^{\rm lin}_{\rm AL}=\frac{e^2T}{2\pi}\,\tGL\,\E ,
\label{eq:lin}
\end{equation}
and the odd-in-$\delta\alpha$ parts of the linear conductivity vanish
identically upon angular integration (the linear-response no-go of
Sec.~\ref{sec:tdgl}(ii) in explicit form).

\subsection{Angular reduction for cubic Lifshitz invariants}
\label{sec:reduction}

All momentum integrals in this paper reduce to one-dimensional radial
master integrals through the following lemma, proved in
\ref{app:reduction} by exact angular averaging. Let $z\equiv Dq^2$, $\alpha_0=z+\Delta$, and let
$C_A(\alpha)$, $C_B(\alpha)$ denote the coefficient functions
multiplying $(\E\cdot\nabla\alpha)^2\,\partial_\q\alpha$ and
$E_iE_j\partial_i\partial_j\alpha\,\partial_\q\alpha$ in
Eqs.~\eqref{eq:ALPGE}--\eqref{eq:ALSHG}, respectively.

\paragraph{(a) $C_{3v}$-type invariant,
$\delta\alpha=\lambda\,q^2(\br\cdot\q)$} The angular average of the
$\mathcal O(\lambda)$ AL integrand produces two radial weights,
\begin{equation}
\big\langle\cdots\big\rangle_i
=P(z)\,(\E\cdot\br)E_i+S(z)\,E^2 b_i,
\qquad
\begin{aligned}
P&=\big[2C_A'z^3+10C_Az^2+4C_Bz\big]/D^2,\\
S&=\big[C_A'z^3+5C_Az^2+2C_B'z^2+6C_Bz\big]/D^2,
\end{aligned}
\label{eq:PS}
\end{equation}
and although $P\neq2S$ pointwise, integration by parts in $z$
[$\int z^kC_A'\,dz=-k\int z^{k-1}C_A\,dz$; all boundary terms vanish]
gives $\int_0^\infty P\,dz=2\int_0^\infty S\,dz$ with
\begin{equation}
\int_0^\infty S\,dz
=\frac{1}{D^2}\int_0^\infty dz\,
\big[2z^2\,C_A(z{+}\Delta)+2z\,C_B(z{+}\Delta)\big] .
\label{eq:Sint}
\end{equation}
The emergent vector structure of the $C_{3v}$ AL response is therefore
exactly
\begin{equation}
\bm F=2\,\E(\E\cdot\br)+\br\,E^2 ,
\label{eq:F}
\end{equation}
i.e., fixed weights $2{:}1$, but only after the radial
integration, a point worth remembering in any generalization (the
kinetic sector of Sec.~\ref{sec:ALkinetic} indeed produces different
weights).

\paragraph{(b) $D_{3h}$ warping invariant,
$\delta\alpha=\lambda\,q_x(q_x^2-3q_y^2)$} The same computation reduces to
\begin{equation}
\big\langle\cdots\big\rangle
=\frac{6z\,\big[C_A z+C_B\big]}{D^2}\;\bm F_w(\E),
\qquad
\bm F_w=(E_x^2-E_y^2,\,-2E_xE_y),
\label{eq:warpred}
\end{equation}
i.e., three times the $C_{3v}$ radial weight of
Eq.~\eqref{eq:Sint}, with the trigonal doublet $\bm F_w$ replacing $\bm
F$. An immediate and nontrivial corollary: the frequency
dependence of the warping-induced AL response is identical to the
$C_{3v}$ one, both invariants are cubic in $q$, so they share the
same radial weights and differ only in their angular content.

\paragraph{(c) MT-type reduction} For an MT integrand of the form
$\intq C_M(\alpha,\beta)\,\partial_i\alpha$ one finds, at
$\mathcal O(\lambda)$ for the $C_{3v}$-type invariant,
\begin{equation}
\intq C_M(\alpha,\beta)\,\partial_i\alpha
\Big|_{\mathcal O(\lambda)}
=\frac{\lambda\,b_i}{4\pi D^2}\int_0^\infty\! dz\,
\mathcal R\big[C_M\big],
\qquad
\mathcal R\big[C\big]\equiv
\Big(\frac{\partial C}{\partial\alpha}\Big)_{\!\beta} z^2+2C z ,
\label{eq:MTred}
\end{equation}
where the $\alpha$-derivative is taken at fixed $\beta$. No integration
by parts is permissible here: $C_M$ depends on $z$ through both
$\alpha=z+\Delta$ and $\beta=z+x\Delta$, so
$\partial_\alpha C_M$ is not the total radial derivative
[$dC_M/dz=\partial_\alpha C_M+\partial_\beta C_M$]. 
For the warping invariant the corresponding angular
averages $\langle\delta\alpha\,q_i\rangle$ and
$\langle\partial_i\delta\alpha\rangle$ vanish identically:
namely, trigonal warping alone produces no MT nonlinearity;
a vector perturbation (strain, or the in-plane $\br$ of the Rashba
problem) is required. This angular selection rule, the MT kernel
retains only the first angular harmonic of the invariant while the
warping term is a pure third harmonic, is insensitive to whether the
invariant resides in the pair spectrum, the relaxation rate, or the
noise vertex, and will apply verbatim to the kinetic sector.

\subsection{$C_{3v}$ results in dimensionless form}
\label{sec:C3vresults}

Applying the reduction (a) to
Eqs.~\eqref{eq:ALPGE}--\eqref{eq:ALSHG} with
$\delta\alpha_{C_{3v}}=\alpha_3q^2(\br\cdot\q)$ and evaluating the
radial integrals in closed form, the AL response reduces to
functions of $\nu=\omega\tGL$ alone, times the singular prefactor
$\tGL^2$. Define
\begin{align}
\mathfrak g(\nu)&=\frac{1}{\nu^2}
+\frac{2}{\nu^4}\Big[2\,\re\big[(2+i\nu)\ln(2+i\nu)\big]-\ln16\Big],
\label{eq:gAL}\\[2pt]
\mathfrak g_c(\nu)&=\frac{1}{2\nu^4}\Big[
-4\ln\!\big(16(1{+}\nu^2)\big)+3\nu^2\ln(1{+}\nu^2)
+4(2{-}\nu^2)\ln(4{+}\nu^2)+4\nu^2\ln4-12\pi\nu
\nonumber\\&\hspace{5.0cm}
+14\nu\arctan\nu+24\nu\arctan(2/\nu)\Big],
\label{eq:gc}\\[2pt]
\mathfrak g_s(\nu)&=\frac{1}{2\nu^4}\Big[
4\pi(\nu^2{-}2)+\nu\big(7\ln(1{+}\nu^2)-12\ln(4{+}\nu^2)+24\ln2\big)
+(8{-}6\nu^2)\arctan\nu
\nonumber\\&\hspace{5.0cm}
+8(2{-}\nu^2)\arctan(2/\nu)\Big],
\label{eq:gs}
\end{align}
with the static values
\begin{equation}
\mathfrak g(0)=\mathfrak g_c(0)=\tfrac{1}{24},
\qquad \mathfrak g_s(0)=0 .
\label{eq:gstatics}
\end{equation}
In terms of these master functions,
\begin{empheq}{align}
\delta\jj^{\rm PGE}_{\rm AL}
&=\frac{e^3T\alpha_3}{\pi D}\,\tGL^{2}\;\mathfrak g(\nu)\;\bm F,
\qquad
\bm F=2\E(\E\cdot\br)+\br E^2,
\label{eq:C3vALPGE}\\
\delta\jj^{\rm SHG}_{\rm AL}
&=\frac{e^3T\alpha_3}{\pi D}\,\tGL^{2}
\big[\mathfrak g_c(\nu)\cos2\omega t
+\mathfrak g_s(\nu)\sin2\omega t\big]\,\bm F .
\label{eq:C3vALSHG}
\end{empheq}
In the static limit
$\delta\jj^{\rm PGE}_{\rm AL}(0)=\delta\jj^{\rm SHG}_{\rm AL}(0)
=\frac{e^3T\alpha_3}{24\pi D}\tGL^2\,\bm F$: the rectified and
second-harmonic amplitudes coincide at dc, as they must, since a
quasistatic quadratic response $\propto E^2(t)=E^2\cos^2\omega t$
splits evenly between the time average and the $\cos2\omega t$
harmonic.

\begin{figure}[t]
\centering
\includegraphics[width=0.49\textwidth]{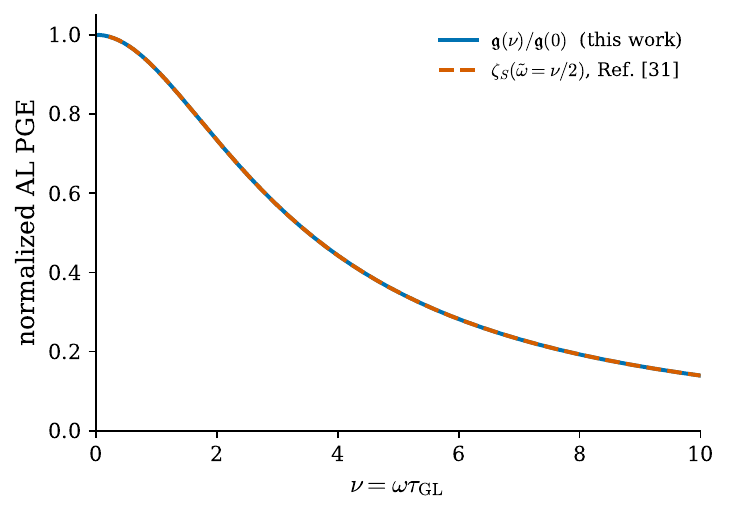}\hfill
\includegraphics[width=0.49\textwidth]{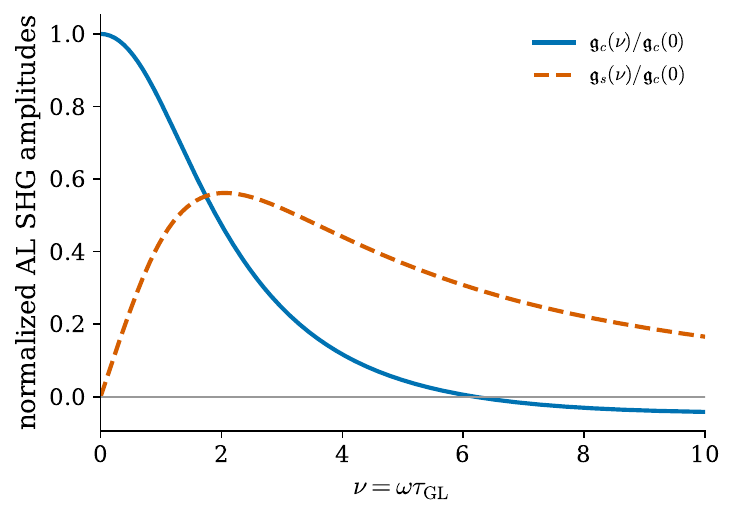}
\caption{Left: normalized AL PGE master function
$\mathfrak g(\nu)/\mathfrak g(0)$ [identical for the $C_{3v}$ invariant
and for both $D_{3h}$ channels], compared with the normalized
photogalvanic response $\zeta_S$ of Parafilo \emph{et al.}
\cite{Parafilo2022} evaluated at $\tilde\omega=\nu/2$: the two frequency
profiles coincide (see text). Right: AL SHG amplitudes
$\mathfrak g_c(\nu)$ (in phase) and $\mathfrak g_s(\nu)$ (quadrature),
normalized by $\mathfrak g_c(0)=1/24$.}
\label{fig:ALfig}
\end{figure}

The frequency profiles are shown in Fig.~\ref{fig:ALfig}. The PGE
master function $\mathfrak g(\nu)$ is Lorentzian-like on the intrinsic
scale $\omega\sim\tGL^{-1}$, with the large-frequency tail
$\mathfrak g\simeq1/\nu^2-2\pi/\nu^3$, which follows from
Eq.~\eqref{eq:gAL} since the bracket vanishes as $\ln\nu/\nu^4$; the
in-phase SHG amplitude $\mathfrak g_c$ changes sign at $\nu=6.19$ and
decays as $-\ln\nu/\nu^2$, while the quadrature amplitude
$\mathfrak g_s$ vanishes at dc, peaks at $\nu=2.06$, and decays as
$\pi/2\nu^2$. Since $\Delta=\tGL^{-1}\propto T-T_c$,
the overall scale of both responses diverges as $\tGL^{2}\propto
(T-T_c)^{-2}$ at fixed $\nu$, which is the fluctuation enhancement
that motivates this work. At fixed laboratory frequency $\omega$
the parameter $\nu=\omega\tGL$ itself grows on approaching $T_c$, so the
$(T-T_c)^{-2}$ growth holds only while $\nu\ll1$, i.e., for
$\epsilon=(T-T_c)/T_c\gg\pi\omega/8T_c$. Closer to $T_c$ the
power-law growth is cut off. The rectified response saturates at the
temperature-independent value $\tGL^2\mathfrak g\to1/\omega^2$ and the
quadrature second harmonic at $\tGL^2\mathfrak g_s\to\pi/2\omega^2$, while
the in-phase second harmonic changes sign at $\omega\tGL=6.19$ and then
grows only logarithmically,
$\tGL^2\mathfrak g_c\to-\ln(\omega\tGL/16)/\omega^2$. The temperature
dependence observed at fixed frequency is therefore the crossover function
$\tGL^2\mathfrak g(\omega\tGL)$ rather than a pure power law, and the
same holds, up to logarithmic factors, for the other master functions
of this paper. The frequency
profile \eqref{eq:gAL} is universal for cubic invariants: it applies
verbatim to both terms of the $D_{3h}$ problem below.

Equation~\eqref{eq:C3vALPGE} can be checked against the literature.
The photogalvanic response of an Ising superconductor was
computed in Ref.~\cite{Parafilo2022} from Langevin-TDGL in a
scalar-potential gauge, with a different truncation and a different
derivation route. Upon the identification $\tilde\omega=\nu/2$ of the
frequency variables, their master function $\zeta_S(\tilde\omega)$ is
proportional to our $\mathfrak g(\nu)$
(Fig.~\ref{fig:ALfig}, left). The agreement of two independent
formalisms on a nontrivial function of frequency validates both
calculations for this contribution: the two expressions agree
although they were obtained in different gauges, with the spatially
inhomogeneous potential $\E=-\nabla \phi(\bm r,t)$ in Ref.~\cite{Parafilo2022}
and the uniform vector potential here $\E=-\partial_t\bm{A}(t)$. The second-harmonic generation of
fluctuating pairs in an Ising superconductor was addressed within the same
scalar-potential formulation in Ref.~\cite{Sonowal2023};
Eqs.~\eqref{eq:D3hALSHG} and \eqref{eq:gc}--\eqref{eq:gs} give the
corresponding AL amplitudes of the warping channel in closed form, with
the trigonal vector structure $\bm F_w$ and the universal profiles
$\mathfrak g_c$, $\mathfrak g_s$. 

\subsection{$D_{3h}$ results: warping and strain}
\label{sec:D3hresults}

For the Ising superconductor the two invariants of
Eq.~\eqref{eq:LI-D3h} contribute additively at linear order. Using
reduction (b) for the warping term [factor $3$, doublet $\bm F_w$] and
reduction (a) for the strain term [identical to $C_{3v}$ with
$\br\to\eps$],
\begin{empheq}{align}
\delta\jj^{\rm PGE}_{\rm AL}
&=\frac{e^3T B_z}{\pi D}\,\tGL^{2}\,\mathfrak g(\nu)\,
\Big[3\kappa\,\bm F_w(\E)+\eta\,\bm F_\eps(\E)\Big],
\qquad
\bm F_\eps=2\E(\E\cdot\eps)+\eps E^2,
\label{eq:D3hALPGE}\\
\delta\jj^{\rm SHG}_{\rm AL}
&=\frac{e^3T B_z}{\pi D}\,\tGL^{2}
\big[\mathfrak g_c(\nu)\cos2\omega t+\mathfrak g_s(\nu)\sin2\omega t\big]
\Big[3\kappa\,\bm F_w(\E)+\eta\,\bm F_\eps(\E)\Big].
\label{eq:D3hALSHG}
\end{empheq}
The frequency content is identical to the $C_{3v}$ case, for the reason
explained below Eq.~\eqref{eq:warpred}. In the static limit the warping
channel gives
$\delta\jj^{\rm PGE}_{\rm AL}(0)+\delta\jj^{\rm SHG}_{\rm AL}(0)
=\frac{e^3T\kappa B_z}{4\pi D}\tGL^2\,\bm F_w$, which reproduces the
full dc nonlinear current of the magnetochiral-anisotropy analysis of
Ref.~\cite{TdM2026a} (with $T\to T_c$), confirming the normalization of
the present finite-frequency treatment against the dc literature.

Note that the strain term contributes to the AL channel on the same
footing as the warping term. For a strain-engineered sample the two
channels are separable by polarization analysis: the trigonal doublet
$\bm F_w$ is locked to the crystallographic axes (it rotates with the
lattice, carrying angular momentum $l=-2$), whereas $\bm F_\eps$ is
locked to the strain axis. Rotating the polarization of the drive at
fixed strain therefore modulates the two contributions with distinct
phases, so that the two contributions can be separated experimentally.

\section{Aslamazov-Larkin channel: kinetic sector}
\label{sec:ALkinetic}

We now turn on the kinetic Lifshitz invariant $\rho(\q)$ and compute its
AL response from the generalized master formula \eqref{eq:genAL}. The
kinetic invariant enters through two inseparable pieces, the
effective rate $\bar\alpha\simeq\alpha-\alpha_0\rho$ in the exponent and the
noise-vertex factor $[1-\rho]$, whose static effects cancel
identically [Eq.~\eqref{eq:gibbs}]; the entire response originates from
their nonequilibrium mismatch along the driven trajectory. Keeping only
one of the two pieces violates the FDT and produces spurious
contributions already at equilibrium; all results below keep the pair
locked.

\subsection{The dc response of the linear drift}
\label{sec:ALdriftdc}

Because the mechanism is novel, we first present the
derivation of the dc limit for the leading new channel, the linear
kinetic drift $\rho=u\,(\br\cdot\q)$, before quoting the systematic
results. Work at dc [$\A=-\E t$, observation time $t=0$] in units
$D=e=1$, and let
\begin{equation}
P\equiv\E\cdot\q,\qquad B_q\equiv\br\cdot\q,\qquad E_b\equiv\E\cdot\br .
\end{equation}
Along the trajectory $\q+2\E s$ the two odd factors are exact
polynomials in the running time,
\begin{equation}
\rho(s)=u\big[B_q+2E_b\,s\big],
\qquad
\alpha_0(s)=\alpha_0+4P s+4E^2 s^2 ,
\end{equation}
so the exponent correction
$X\equiv\int_{t'}^{0}\bar\alpha\,ds+\alpha_0 t'$ and the noise factor
$[1-\rho(t')]$ can be expanded to $\mathcal O(uE^2)$ in closed form.
Using $\int_{-\infty}^0 t'^n e^{2\alpha_0t'}dt'
=(-1)^n n!/(2\alpha_0)^{n+1}$, the $\mathcal O(uE^2)$ part of the time
integral of Eq.~\eqref{eq:genAL} collects into
\begin{equation}
K_{uE^2}
=\left(\frac{-2B_qE^2-PE_b}{\alpha_0^4}
+\frac{12\,B_qP^2}{\alpha_0^5}\right)u
+\big(\text{$\q$-even terms}\big),
\label{eq:KuE2}
\end{equation}
where the $\q$-even terms drop after angular averaging together with
the vertex $2q_i$.
Attaching the vertex $2q_i$ and averaging over angles with the exact 2D
moments
\begin{equation}
\langle2q_iB_q\rangle=z\,b_i,\qquad
\langle2q_iP\rangle=z\,E_i,\qquad
\langle2q_iB_qP^2\rangle=\tfrac{z^2}{4}\big(2E_bE_i+E^2b_i\big),
\qquad z=q^2,
\end{equation}
and evaluating the radial integrals
$\int_0^\infty z\,dz/\alpha_0^4=\tfrac{1}{6}\tGL^2$ and
$\int_0^\infty z^2dz/\alpha_0^5=\tfrac{1}{12}\tGL^2$, one finds
\begin{equation}
\delta\jj^{u}_{\rm AL}
=\frac{e^3Tu}{\pi}\Big[
\tfrac16\big({-}2E^2\br-E_b\E\big)
+\tfrac14\big(2E_b\E+E^2\br\big)\Big]\tGL^2
=\frac{e^3T\,u}{12\pi}\,\tGL^2\,
\big[\,4\,\E(\E\cdot\br)-\br\,E^2\,\big].
\label{eq:ALdrift-dc}
\end{equation}
The derivation shows the mechanism. A fluctuating
pair with momentum along $+\hat{\bm u}$ relaxes faster and is
pumped harder than its $-\hat{\bm u}$ partner; in equilibrium the two
effects cancel identically (FDT), but the field-driven trajectory samples
the rate and the noise at different momenta, and the mismatch
rectifies.

Three properties of Eq.~\eqref{eq:ALdrift-dc} follow. The
linear drift contributes at the same leading order $\tGL^2$ as
the thermodynamic cubic invariant [Eq.~\eqref{eq:C3vALPGE}]. Since
$[u]=\mathrm{length}$, no factor of $D$ appears, and microscopically
$u\,\xi^{-1}\sim\sqrt{\epsilon}$ matches the thermodynamic channel's
$\alpha_3 q^3/\alpha\sim\sqrt{\epsilon}$ at the fluctuation momentum
scale, so the two channels are parametrically comparable. The
polarization weights differ, $(4{:}{-}1)$ for the drift channel versus
$(2{:}1)$ for the thermodynamic one, so the two are distinguishable by
polarization analysis; for instance, for $\E\perp\br$ the two
channels drive currents in opposite directions along $\br$.
The result also survives in the $D_{3h}$ problem with strain, where
the same-representation kinetic drift $\rho=\upsilon B_z(\eps\cdot\q)$, 
non-removable, in contrast to its thermodynamic counterpart, 
gives Eq.~\eqref{eq:ALdrift-dc} with $u\to\upsilon B_z$, $\br\to\eps$:
a leading-order, strain-locked AL contribution.

\subsection{Systematic dc results for all kinetic channels}
\label{sec:ALdriftall}

Carrying out the same program for the remaining invariants of
Eqs.~\eqref{eq:rho-C3v}--\eqref{eq:rho-D3h} (the systematic expansion
was performed with computer algebra and cross-checked as described in
\ref{app:verification}) yields the complete dc catalogue of the AL
channel, thermodynamic and kinetic:
\begin{center}
\begin{tabular}{llll}
\hline\hline
channel & invariant & $\delta\jj_{\rm AL}(\omega{=}0)$ & singularity\\
\hline
thermodynamic cubic & $\delta\alpha=\alpha_3q^2(\br\cdot\q)$ &
$\dfrac{e^3T\alpha_3}{12\pi D}\,\tGL^2\,[2\E(\E\cdot\br)+\br E^2]$ &
$\tGL^2$\\[8pt]
kinetic linear drift & $\rho=u\,(\br\cdot\q)$ &
$\dfrac{e^3T\,u}{12\pi}\,\tGL^2\,[4\E(\E\cdot\br)-\br E^2]$ &
$\tGL^2$\\[8pt]
kinetic cubic drift & $\rho=u_3q^2(\br\cdot\q)$ &
$\dfrac{e^3T\,u_3}{12\pi D}\,\tGL\,[10\E(\E\cdot\br)-\br E^2]$ &
$\tGL$\\[8pt]
thermodynamic warping & $\delta\alpha=\kappa B_zq_x(q_x^2-3q_y^2)$ &
$\dfrac{e^3T\kappa B_z}{4\pi D}\,\tGL^2\,\bm F_w(\E)$ &
$\tGL^2$\\[8pt]
kinetic warping & $\rho=\tilde\kappa B_zq_x(q_x^2-3q_y^2)$ &
$\dfrac{e^3T\,\tilde\kappa B_z}{4\pi D}\,\tGL\,\bm F_w(\E)$ &
$\tGL$\\[4pt]
\hline\hline
\end{tabular}
\end{center}
\smallskip
\noindent
[Strain channels of the $D_{3h}$ problem follow from the first three
rows with $\br\to\eps$ and $\alpha_3\to\eta B_z$, $u\to\upsilon B_z$,
$u_3\to\tilde\eta B_z$.] The pattern is the following. Kinetic invariants
that are structurally identical to a thermodynamic invariant
(cubic drift, kinetic warping) produce the same vector structures with
one power of $\tGL$ less, they are subleading near $T_c$, suppressed
by $\sim(T-T_c)/T_c$ relative to their thermodynamic partners. The
linear drift, which has no non-removable thermodynamic counterpart, is
the exception: it enters at the full $\tGL^2$ singularity and
is a new leading-order channel. For unstrained
$D_{3h}$, where symmetry forbids any linear invariant, the kinetic
sector merely renormalizes the magnitude of the (unique) $\bm F_w$
response at relative order $T-T_c$: the published dc conclusions of
Ref.~\cite{TdM2026a} are robust. For $C_{3v}$, and for strained
$D_{3h}$, the drift channel is a same-order additive contribution with
distinct polarization weights.

\subsection{Finite frequency: drift master functions}
\label{sec:ALdriftomega}

At finite drive frequency the linear-drift channel acquires its own
frequency profile, different from the thermodynamic
$\mathfrak g(\nu)$. Repeating the expansion of
Sec.~\ref{sec:ALtime} with the drift factors included [the $\rho$
insertions oscillate at $\pm\omega$, like the $\delta\alpha$
gradient terms, but enter with different time weights], the rectified
response reads
\begin{equation}
\delta\jj^{\rm PGE}_{u}
=\frac{e^3T u}{\pi}
\int_0^\infty\!\!dz\,\Big[
c_1\, z\,\E(\E\cdot\br)
+c_2\, z\,\br E^2
+c_3\,\frac{z^2}{4}\big(2\E(\E\cdot\br)+\br E^2\big)\Big],
\label{eq:ALdriftPGE}
\end{equation}
with the radial coefficient functions [$\alpha\equiv\alpha_0(z)=z+\Delta$]
\begin{equation}
c_1=-\frac{2(4\alpha^2-\omega^2)}{\alpha^2(4\alpha^2+\omega^2)^2},
\qquad
c_2=-\frac{16}{(4\alpha^2+\omega^2)^2},
\qquad
c_3=\frac{32(12\alpha^2-\omega^2)}{\alpha(4\alpha^2+\omega^2)^3}.
\label{eq:c123}
\end{equation}
Collecting the two vector structures,
\begin{equation}
\delta\jj^{\rm PGE}_{u}
=\frac{e^3Tu}{\pi}\,\tGL^{2}\,
\Big[\,\mathcal U_1(\nu)\,\E(\E\cdot\br)
+\mathcal U_2(\nu)\,\br E^2\Big],
\qquad
\mathcal U_1(0)=\tfrac16,\quad
\mathcal U_2(0)=-\tfrac1{24},
\label{eq:Ufuncs}
\end{equation}
where $\mathcal U_1=\int dz\,[c_1z+\tfrac12 c_3z^2]$ and
$\mathcal U_2=\int dz\,[c_2z+\tfrac14 c_3z^2]$ (evaluated in units
$\Delta=1$, $\alpha=z+1$). At $\nu\to0$ the rectified amplitudes
approach one half of the dc weights of
Eq.~\eqref{eq:ALdrift-dc}, $[\tfrac16,-\tfrac1{24}]
=\tfrac12[\tfrac13,-\tfrac1{12}]$; the other half resides in the
in-phase second harmonic, in accord with the sum rule
\eqref{eq:sumrule} below. The master functions are
plotted in Fig.~\ref{fig:ALdrift} (left) against the thermodynamic
profile $\mathfrak g(\nu)$. Both weights retain their dc signs at all
frequencies ($\mathcal U_1>0$, $\mathcal U_2<0$), and their ratio
stays close to the dc value, $\mathcal U_2/\mathcal U_1$ drifting only
weakly from $-\tfrac14$ at $\nu=0$ to $\simeq-0.29$ at $\nu\sim5$: as
in the thermodynamic channel, the polarization anisotropy of the PGE
is nearly frequency independent. The experimental discriminator between the two
channels is therefore the value of the anisotropy --- weights
$(4{:}{-}1)$, with the $\br E^2$ component reversed, versus
$(2{:}1)$ --- rather than its frequency dependence. The corresponding
$2\omega$ (SHG) amplitudes follow from the same computation; their
$\omega\to0$ limits are fixed by the exact sum rule
\begin{equation}
\lim_{\omega\to0}\big[\text{PGE}+\text{in-phase SHG}\big]
=\text{full dc response},
\label{eq:sumrule}
\end{equation}
which we verified term by term for every channel in this paper, at
dc the quadratic response to $\E\cos\omega t$ splits between the time
average and the $\cos2\omega t$ harmonic, and Eq.~\eqref{eq:sumrule}
is an independent check of the separately computed harmonics.

\begin{figure}[t]
\centering
\includegraphics[width=0.98\textwidth]{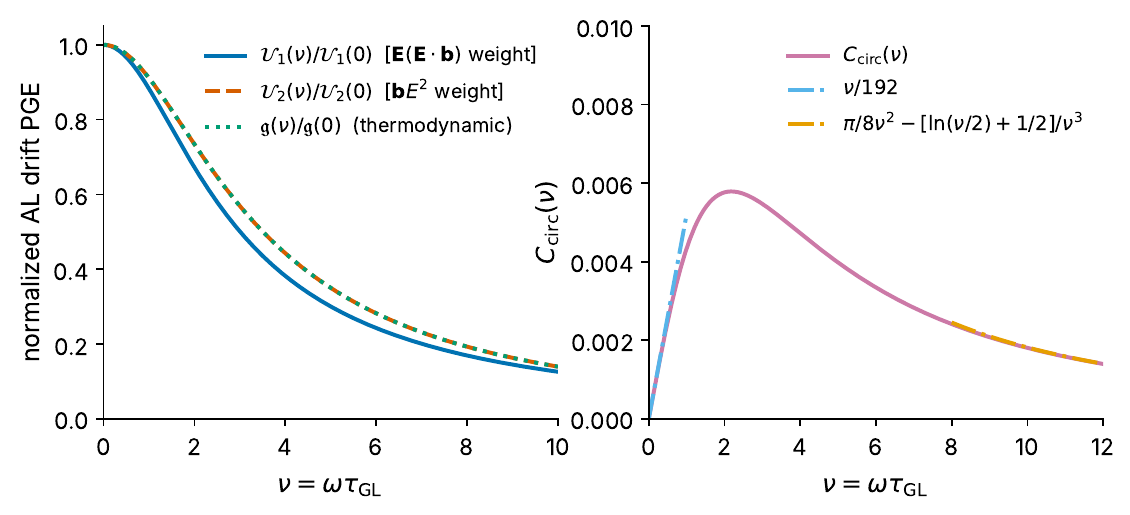}
\caption{Left: normalized PGE master functions of the AL linear-drift
(kinetic) channel, Eq.~\eqref{eq:Ufuncs}: the $\E(\E\cdot\br)$ weight
$\mathcal U_1$ and the $\br E^2$ weight $\mathcal U_2$, compared with
the thermodynamic profile $\mathfrak g(\nu)$ (dotted). Both weights
retain their dc signs at all frequencies, with the anisotropy ratio
$\mathcal U_2/\mathcal U_1$ remaining close to its dc value
$-\tfrac14$. Right: the helicity-odd
(circular) AL master function $C_{\rm circ}(\nu)$ of
Sec.~\ref{sec:PVHE}, which vanishes linearly at dc [dash-dotted line:
the small-$\nu$ asymptote $\nu/192$], peaks at
$\omega\simeq2.2\,\tGL^{-1}$, and decays as $\pi/8\nu^2$.}
\label{fig:ALdrift}
\end{figure}

\section{Maki-Thompson channel}
\label{sec:MT}

\subsection{Finite-frequency kernels: thermodynamic sector}
\label{sec:MTthermo}

We evaluate the master formula \eqref{eq:genMT} at $\rho=0$, to second
order in the field and to linear order in $\delta\alpha$. With the gauge
\eqref{eq:gauge} the accelerated momentum in the Langevin factor is
\begin{equation}
\alpha\big(\q-2e\A(t'')\big)
=\alpha\Big(\q+\tfrac{2e}{\omega}\,\E\sin\omega t''\Big)
\simeq\alpha_\q+\frac{2}{\omega}\,\sin(\omega t'')\,g_1,
\qquad
g_1\equiv e\,\E\cdot\nabla\alpha_\q ,
\label{eq:MTexpand}
\end{equation}
so the inner exponent acquires the correction
\begin{equation}
-2\int_{t_2}^{t_1}\!dt''\,\alpha\big(\q-2e\A(t'')\big)
=-2\alpha_\q(t_1-t_2)
+\frac{4g_1}{\omega^2}\big[\cos\omega t_1-\cos\omega t_2\big].
\label{eq:MTinner}
\end{equation}
The Cooperon pole is evaluated at the accelerated momentum
$\q-e\A(t')-e\A(2t_1-t')=\q+\tfrac{2e}{\omega}\E\sin\omega t_1\cos\omega(t'-t_1)$,
so the outer exponent acquires the correction
\begin{equation}
-2\int_{t_1}^{t}\!dt'\,\beta\big(\q-e\A(t')-e\A(2t_1-t')\big)
=-2\beta_\q(t-t_1)-\frac{4h_1}{\omega^2}\,\sin\omega t_1\,\sin\omega(t-t_1),
\qquad h_1\equiv e\,\E\cdot\nabla\beta_\q ,
\label{eq:MTdrift}
\end{equation}
to first order in the field; the second-order term of this expansion is
independent of the direction of $\q$ and drops out of the
$\mathcal O(\delta\alpha E^2)$ response after angular averaging.
Expanding the
exponentials to the required order in $g_1$ and $h_1$, performing
the $t_2$ integral, and then integrating $t_1$ against the Cooperon weight
and the vertex $\E\cos\omega(2t_1-t)$
generates, after harmonic decomposition, the complete quadratic MT
response:
\begin{empheq}{align}
\jj^{(2)}_{\rm MT}(\omega)
=16e^2TD\,\E\intq
\Big\{\Big[&
-\frac{2(\alpha-\beta)}{\alpha(4\alpha^2+\omega^2)(4\beta^2+\omega^2)}
+\frac{2(3\alpha+\beta)\cos 2\omega t}
{\alpha(4\alpha^2+\omega^2)(4\beta^2+9\omega^2)}
\nonumber\\
&-\frac{(4\alpha\beta-3\omega^2)\sin 2\omega t}
{\omega\,\alpha(4\alpha^2+\omega^2)(4\beta^2+9\omega^2)}
\Big]\,g_1
\nonumber\\[2pt]
+\Big[&
\frac{1}{4\alpha^2\beta(\beta^2+\omega^2)}
-\frac{3\beta\cos2\omega t}{4\alpha^2(\beta^2+\omega^2)(\beta^2+4\omega^2)}
\nonumber\\
&+\frac{(\beta^2-2\omega^2)\sin2\omega t}
{4\alpha^2\omega(\beta^2+\omega^2)(\beta^2+4\omega^2)}
\Big]\,h_1\,\delta\alpha_\q\Big\} .
\label{eq:MTkernels}
\end{empheq}
The first bracket collects the terms in which the field enters through
the pair side, with $g_1$ containing the gradient of the full spectrum
$\alpha_\q=\alpha_0+\delta\alpha$; the second bracket is the Cooperon
drift, in which the field enters through $h_1$ and the nonreciprocity
through the static shift $-\delta\alpha_\q/2\alpha_\q^2$ of the pair
propagator. The two brackets are of the same order and must be kept
together. The pair relaxation pole enters as $4\alpha^2+\omega^2$ through the
two Langevin propagators. The Cooperon enters as $4\beta^2+\omega^2$ in
the rectified part and as $4\beta^2+9\omega^2$ in the second harmonic,
because the vertex oscillates at $2t_1-t$ and the retarded-advanced
Cooperon pair carries three frequency quanta in the $2\omega$ channel;
the drift term adds the single-Cooperon factors $\beta^2+\omega^2$
and $\beta^2+4\omega^2$. The quadrature amplitudes carry an explicit
$1/\omega$; the residue of this pole is a total momentum derivative
and vanishes upon momentum integration, as shown in
Sec.~\ref{sec:MTsecular}. For
reference, the same expansion at linear order in the field reproduces
the classical anomalous MT paraconductivity
\cite{Thompson1970,LarkinVarlamov2005}: at $\omega\to0$,
\begin{equation}
\jj^{\rm lin}_{\rm MT}
=\frac{e^2T}{\pi}\,
\frac{\ln\big(\tau_\phi/\tGL\big)}{\tGL^{-1}-\tau_\phi^{-1}}\,\E ,
\label{eq:linMT}
\end{equation}
the quantum-interference counterpart of the AL result
\eqref{eq:lin}.

Applying the MT reduction \eqref{eq:MTred} to the first bracket of
Eq.~\eqref{eq:MTkernels}, and the angular average
$\langle h_1\delta\alpha_\q\rangle=\lambda z^2 e\,(\E\cdot\br)$ (units
$D=1$) to the second, converts
Eq.~\eqref{eq:MTkernels} into dimensionless master functions. Define,
in units $\Delta=1$ [$\alpha=z+1$, $\beta=z+x$],
\begin{align}
&m_{\rm dc}(\nu,x)=\int_0^\infty\! dz\;\Big\{
\mathcal R\Big[\frac{-2(\alpha-\beta)}
{\alpha(4\alpha^2+\nu^2)(4\beta^2+\nu^2)}\Big]
+\frac{z^2}{4\alpha^2\beta(\beta^2+\nu^2)}\Big\},
\nonumber\\
&m_c(\nu,x)=\int_0^\infty\! dz\;\Big\{
\mathcal R\Big[\frac{2(3\alpha+\beta)}
{\alpha(4\alpha^2+\nu^2)(4\beta^2+9\nu^2)}\Big]
-\frac{3\beta z^2}{4\alpha^2(\beta^2+\nu^2)(\beta^2+4\nu^2)}\Big\},
\nonumber\\
&m_s(\nu,x)=\int_0^\infty\! dz\;\Big\{
\mathcal R\Big[\frac{-(4\alpha\beta-3\nu^2)}
{\nu\,\alpha(4\alpha^2+\nu^2)(4\beta^2+9\nu^2)}\Big]
+\frac{z^2(\beta^2-2\nu^2)}{4\alpha^2\nu(\beta^2+\nu^2)(\beta^2+4\nu^2)}\Big\},
\nonumber\\
&m_{\rm circ}(\nu,x)=\int_0^\infty\! dz\;\Big\{
\mathcal R\Big[\frac{4\alpha\beta+\nu^2}
{\nu\,\alpha(4\alpha^2+\nu^2)(4\beta^2+\nu^2)}\Big]
-\frac{z^2}{4\alpha^2\nu(\beta^2+\nu^2)}\Big\},
\label{eq:mfuncs}
\end{align}
where the last function anticipates the circular response of
Sec.~\ref{sec:PVHE}, and in each case the second term is the
Cooperon-drift contribution. The static limits of the first two
functions coincide and evaluate in closed form,
\begin{equation}
m_{\rm dc}(0,x)=m_{c}(0,x)=\frac{f(x)}{8},
\qquad
f(x)=\frac{2x^{2}\ln x-5x^{2}+4x\ln x+4x+1}{2\,(x-1)^{4}},
\label{eq:mdc0}
\end{equation}
so that the rectified and in-phase second-harmonic amplitudes split
evenly at $\omega\to0$, as they must for a quasistatic quadratic
response [cf.\ the discussion below Eq.~\eqref{eq:C3vALSHG}], and their
sum,
\begin{equation}
m_{\rm dc}(0,x)+m_c(0,x)=\frac14\,f(x),
\label{eq:dcsum}
\end{equation}
is the dephasing function of the dc magnetochiral MT
kernel $\propto1/(\beta\alpha^3)$: the finite-frequency rectified-plus-in-phase
response reduces at $\omega\to0$ to the full dc nonlinear current,
Eq.~\eqref{eq:sumrule}. The drift term is essential for this result:
without it the static split is uneven and the dc total acquires the
additional kernel $\propto1/(\alpha^2\beta^2)$ that was retained in the
dc analysis of Refs.~\cite{TdM2026a,TdM2026b}. The Cooperon drift cancels
this term through the identity
$g_1/4\alpha^2\beta^2+h_1/2\alpha\beta^3
=-e\,\E\cdot\partial_\q(1/4\alpha\beta^2)$, so that $f(x)$ replaces the
dephasing function quoted there. The limiting
behaviors of $f$ are
\begin{equation}
f(0)=\tfrac12,\qquad f(1)=\tfrac1{12},\qquad
f(x\gg1)\simeq\frac{\ln x}{x^{2}} .
\label{eq:ffull-limits}
\end{equation}
The MT dc total is therefore
\begin{equation}
\delta\jj_{\rm MT}(\omega{=}0)=\frac{e^3T\alpha_3}{\pi D}\,\tGL^2\,f(x)\,\E(\E\cdot\br),
\label{eq:MTdctotal}
\end{equation}
which, compared with the AL dc total $(e^3T\alpha_3/12\pi D)\,\tGL^2\,\bm F$
of Sec.~\ref{sec:C3vresults}, carries the
$\E(\E\cdot\br)$ weight $6f(x)$ relative to the AL one: three times
larger at weak pair breaking and one half at $x=1$. The logarithm
$\ln(\tau_\phi/\tGL)$ of the linear MT conductivity \eqref{eq:linMT} has
no counterpart in the nonlinear response: the kernel $1/(\alpha\beta)$ of
the linear response, whose momentum integral is logarithmic in $x$, is
replaced in Eq.~\eqref{eq:MTkernels} by kernels with higher powers of
$\alpha$, whose integrals remain finite as $x\to0$.

\subsection{Results for the two point groups}
\label{sec:MTresults}

\begin{figure}[t]
\centering
\includegraphics[width=0.49\textwidth]{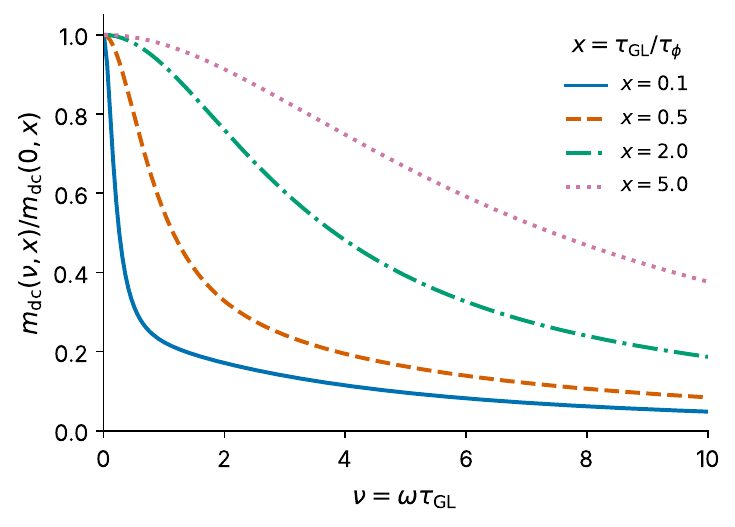}\hfill
\includegraphics[width=0.49\textwidth]{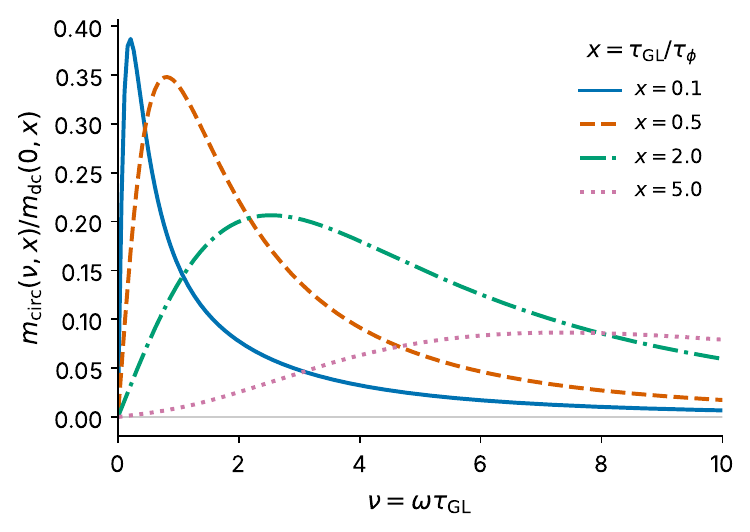}
\caption{Left: normalized MT PGE master function
$m_{\rm dc}(\nu,x)/m_{\rm dc}(0,x)$ for several dephasing ratios
$x=\tGL/\tau_\phi$, with $m_{\rm dc}(0,x)=f(x)/8$. Right: the helicity-odd (circular) MT master
function $m_{\rm circ}(\nu,x)$ of Sec.~\ref{sec:PVHE}, normalized by
$m_{\rm dc}(0,x)$; it vanishes linearly at $\nu\to0$ with the slope
$s_{\rm circ}(x)$ of Eq.~\eqref{eq:mcirc-res} and peaks at
$\omega\sim\tau_\phi^{-1}$ for weak pair breaking.}
\label{fig:MTPGEfig}
\end{figure}

\begin{figure}[t]
\centering
\includegraphics[width=0.9\textwidth]{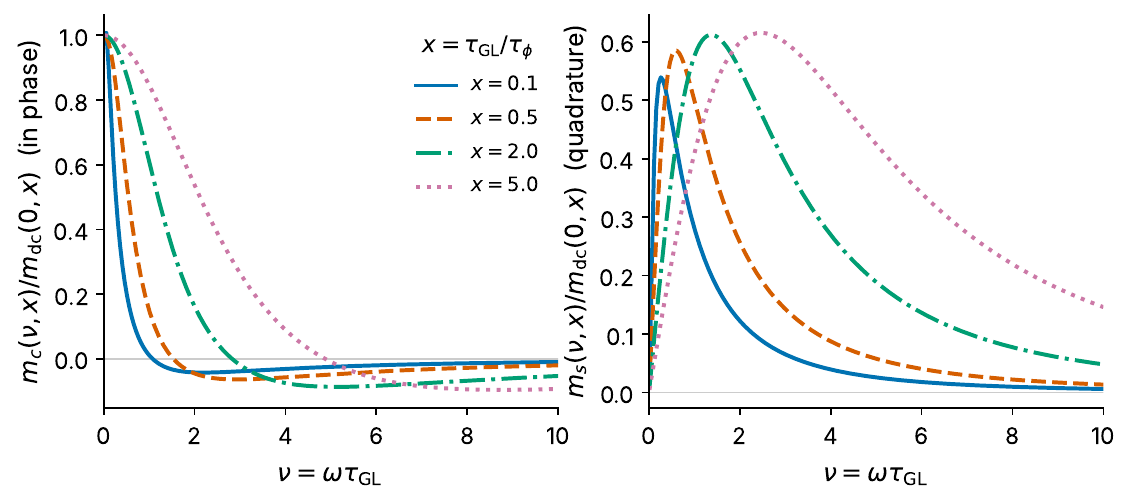}
\caption{MT SHG amplitudes, normalized by the static PGE magnitude
$m_{\rm dc}(0,x)=f(x)/8$, for
several dephasing ratios: in-phase $m_c$ (left), which changes sign at
$\nu_0(x)$, and quadrature
$m_s$ (right; the amplitude vanishes linearly at small $\nu$,
Eq.~\eqref{eq:slin}).}
\label{fig:MTSHGfig}
\end{figure}

For the $C_{3v}$ invariant, the reduction gives
\begin{empheq}{align}
\delta\jj^{\rm PGE}_{\rm MT}
&=\frac{4e^3T\alpha_3}{\pi D}\,\tGL^{2}\;m_{\rm dc}(\nu,x)\,
\E\,(\E\cdot\br),
\label{eq:C3vMTPGE}\\
\delta\jj^{\rm SHG}_{\rm MT}
&=\frac{4e^3T\alpha_3}{\pi D}\,\tGL^{2}
\big[m_c(\nu,x)\cos2\omega t+m_s(\nu,x)\sin2\omega t\big]\,
\E\,(\E\cdot\br),
\label{eq:C3vMTSHG}
\end{empheq}
where $(\E\cdot\br)=[\E\times\bm B]_z$: in contrast to the AL channel,
MT produces a pure $\E(\E\cdot\br)$ structure, no
$\br E^2$ term. For the $D_{3h}$ problem the warping invariant drops
out identically by the angular selection rule, and the MT nonlinearity is a pure strain effect:
\begin{empheq}{align}
\delta\jj^{\rm PGE}_{\rm MT}
&=\frac{4e^3T\eta B_z}{\pi D}\,\tGL^{2}\,m_{\rm dc}(\nu,x)\,
\E\,(\E\cdot\eps),
\label{eq:D3hMTPGE}\\
\delta\jj^{\rm SHG}_{\rm MT}
&=\frac{4e^3T\eta B_z}{\pi D}\,\tGL^{2}
\big[m_c\cos2\omega t+m_s\sin2\omega t\big]\,\E\,(\E\cdot\eps).
\label{eq:D3hMTSHG}
\end{empheq}
The master functions are plotted in
Figs.~\ref{fig:MTPGEfig}--\ref{fig:MTSHGfig} for several values of the
dephasing ratio. Unlike the AL channel, whose profile is fixed by
$\nu$ alone, the MT response depends on both the intrinsic scale
$\omega\sim\tGL^{-1}$ and the dephasing scale $\omega\sim\tau_\phi^{-1}$.
The rectified amplitude $m_{\rm dc}$ is positive and decays
monotonically. The in-phase second-harmonic amplitude $m_c$ starts at the
same static value $f(x)/8$, changes sign at a dephasing-dependent
frequency $\nu_0(x)$, and remains negative at higher frequency.
The quadrature amplitude vanishes linearly at low frequency,
\begin{align}
m_s(\nu\to0,x)&=\nu\,s_{\rm lin}(x),\qquad s_{\rm lin}(1)=\tfrac1{80},
\nonumber\\
s_{\rm lin}(x)&=\frac{x^4-46x^3-18x^2+62x+1+12x(x^2+6x+2)\ln x}{48\,x\,(x-1)^5},
\label{eq:slin}
\end{align}
with $s_{\rm lin}\simeq-1/48x$ for $x\ll1$ (the sign of the onset
reverses at $x\simeq0.04$) and $s_{\rm lin}\simeq1/48x^2$ for $x\gg1$. The
measurement of the frequency profile at several temperatures, and in
particular of the zero crossing $\nu_0(x)$ of the in-phase second
harmonic, therefore provides a direct experimental determination of
$\tau_\phi$, in the same spirit as the classical MT analysis of the
linear magnetoconductivity, but now in a contact-free optical setting.

\subsection{Low-frequency limit and cancellation of the secular terms}
\label{sec:MTsecular}

A strictly dc evaluation of the master formula \eqref{eq:genMT} in the
gauge $\A=-\E t$ produces, alongside the stationary kernels, a term
growing linearly with the observation time, the ``secular term'', whose
$\omega\to0$ counterpart is the $1/\omega$ pole of the quadrature
harmonic, $\sin(2\omega t)/\omega\to2t$. Both features are artifacts of
truncating the field dependence of the Cooperon pole. At the level of
the per-momentum kernels, the pair-side drift $g_1$ and the Cooperon
drift $h_1$ produce secular pieces that combine into a total momentum
gradient,
\begin{equation}
\frac{g_1}{\alpha^2\beta}+\frac{h_1}{\alpha\beta^2}
=-\,e\,\E\cdot\partial_\q\frac{1}{\alpha\beta},
\qquad h_1\equiv e\,\E\cdot\nabla\beta_\q ,
\label{eq:secular-cancel}
\end{equation}
which vanishes identically under $\intq$ [the boundary term dies as
$1/(\alpha\beta)\sim q^{-4}$, and $\alpha,\beta>0$ above $T_c$].
The same cancellation operates at finite frequency. The residues of
the $1/\nu$ poles of the quadrature and helicity-odd integrands of
Eq.~\eqref{eq:mfuncs} are total derivatives in the radial variable,
\begin{equation}
\lim_{\nu\to0}\nu\,\big[m_s\text{ integrand}\big]
=-\frac14\,\frac{d}{dz}\,\frac{z^2}{\alpha^2\beta},
\qquad
\lim_{\nu\to0}\nu\,\big[m_{\rm circ}\text{ integrand}\big]
=+\frac14\,\frac{d}{dz}\,\frac{z^2}{\alpha^2\beta},
\label{eq:residues}
\end{equation}
and integrate to zero because $z^2/\alpha^2\beta\to1/z$ at large $z$ and
vanishes at $z=0$. Without the drift term the residues,
$z(z-\alpha)/2\alpha^3\beta$ and its negative, are not total derivatives
and the momentum-integrated quadrature amplitudes diverge as $1/\nu$;
with it, $m_s$ and $m_{\rm circ}$ are finite at all frequencies and
vanish linearly at $\nu\to0$ [Eqs.~\eqref{eq:slin} and
\eqref{eq:mcirc-res}]. An equivalent statement follows from
manifest stationarity: shifting $\q\to\q-2e\E t$ removes all reference
to absolute time from the dc master formula, order by order in the
field. Therefore the dc steady state exists and the measured
quantities, the rectified current and the two SHG quadratures, are
finite. The result does not depend on the dc prescription (strict dc
gauge, adiabatic switching, or the $\omega\to0$ limit of a periodic
drive). A detailed analysis of the two channels is given in
Ref.~\cite{TdM2026a}.

\subsection{Kinetic sector of the MT channel}
\label{sec:MTkinetic}

The kinetic Lifshitz invariant enters the MT master formula
\eqref{eq:genMT} through the same locked pair as in the AL channel: the
noise-vertex factor $[1-\rho(\q-2e\A(t_2))]$ and the rate shift
$-\alpha_0\rho$ inside the Langevin exponent. The Cooperon drift
\eqref{eq:MTdrift} does not contribute in this sector: it multiplies the
$\mathcal O(\rho E^0)$ part of the pair side, which vanishes identically
by the FDT lock [Eq.~\eqref{eq:gibbs}]. Expanding at dc to second
order in the field and to linear order in $\rho$, and applying the
angular reduction, one finds the following catalogue [$m_1$ and $m_3$
plotted in Fig.~\ref{fig:MTdriftfig}]:
\begin{center}
\begin{tabular}{llll}
\hline\hline
channel & invariant & $\delta\jj_{\rm MT}(\omega{=}0)$ & singularity\\
\hline
thermodynamic cubic & $\alpha_3q^2(\br\cdot\q)$ &
$\dfrac{e^3T\alpha_3}{\pi D}\,\tGL^2\, f(x)\,
\E(\E\cdot\br)$ &
$\tGL^2$\\[8pt]
kinetic linear drift & $u\,(\br\cdot\q)$ &
$\dfrac{e^3T\,u}{2\pi}\,\tGL^2\; m_1(x)\,\E(\E\cdot\br)$ &
$\tGL^2$\\[8pt]
kinetic cubic drift & $u_3\,q^2(\br\cdot\q)$ &
$\dfrac{e^3T\,u_3}{2\pi D}\,\tGL\; m_3(x)\,\E(\E\cdot\br)$ &
$\tGL$\\[8pt]
kinetic warping ($D_{3h}$) & $\tilde\kappa B_zq_x(q_x^2-3q_y^2)$ &
$0$ \quad(identically) & ---\\[2pt]
\hline\hline
\end{tabular}
\end{center}
\smallskip
\noindent with the dephasing functions
\begin{equation}
m_1(x)=\frac{1-x^2+2x\ln x}{(1-x)^3},
\qquad
m_3(x)=\frac{1-4x+3x^2-2x^2\ln x}{(1-x)^3},
\label{eq:m1m3}
\end{equation}
normalized to $m_1(0)=m_3(0)=1$, with $m_1(1)=\tfrac13$,
$m_3(1)=\tfrac23$, and large-$x$ tails $m_1\simeq1/x$,
$m_3\simeq(2\ln x-3)/x$. As in the AL channel, strain rows for the
$D_{3h}$ problem follow with $\br\to\eps$ and the couplings
$u\to\upsilon B_z$, $u_3\to\tilde\eta B_z$.

Two conclusions follow. The hierarchy of $\tGL$ powers is the same
as in the AL channel: the linear kinetic drift enters at the full
$\tGL^2$ singularity, on the same footing as the thermodynamic invariant,
while the cubic drift carries one power of $\tGL$ less and the kinetic
warping invariant vanishes by the angular selection rule.~\footnote{The
dimensional restoration follows from the rescaling of the radial
variable by $\Delta$: the dimensionless couplings are
$u\sqrt{\Delta/D}$ for the linear and $u_3(\Delta/D)^{3/2}$ for the
cubic drift, against $\alpha_3\Delta^{1/2}D^{-3/2}$ for the thermodynamic
invariant, and the current unit is $eT\sqrt{\Delta/D}$ in both
channels.} Both channels therefore mix the thermodynamic and the linear
kinetic sector at leading order; in the MT channel the ratio of the
kinetic to the thermodynamic $\E(\E\cdot\br)$ weight is
$uD\,m_1(x)/2\alpha_3f(x)$, of order unity when $u\sim\alpha_3/D$. The two
sectors are nevertheless distinguishable. In the AL channel they carry
different polarization weights, $(4{:}{-}1)$ versus $(2{:}1)$; in the MT
channel they share the pure $\E(\E\cdot\br)$ structure but carry
different dephasing functions, $m_1(x)$ against $f(x)$, and different
frequency profiles (Sec.~\ref{sec:MTkineticomega}); and the helicity-odd
AL response of Sec.~\ref{sec:ALcirc} exists only in the kinetic sector.
Measuring both channels, separable by their dephasing dependences and
polarization structures, thus disentangles the nonreciprocity of the pair
spectrum from that of the pair kinetics.

\subsection{Kinetic sector at finite frequency}
\label{sec:MTkineticomega}

\begin{figure}[t]
\centering
\includegraphics[width=0.98\textwidth]{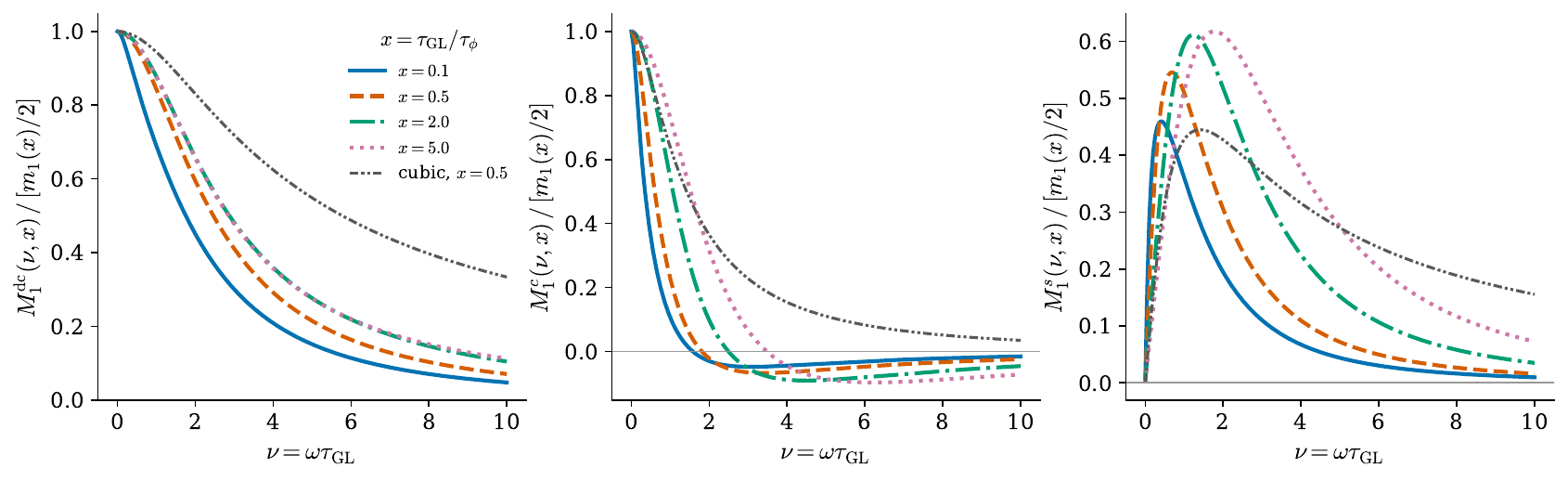}
\caption{Finite-frequency master functions of the kinetic (linear
drift) MT channel, normalized by the static value $m_1(x)/2$, for
several dephasing ratios [gray: cubic drift at $x=0.5$, normalized by
$m_3(0.5)/2$]. Left: rectified amplitude $M_1^{\rm dc}$; middle:
in-phase SHG amplitude $M_1^{c}$; right: quadrature amplitude
$M_1^{s}$, with the linear onset of Eq.~\eqref{eq:s1}; compare the
thermodynamic MT channel of Fig.~\ref{fig:MTSHGfig}.}
\label{fig:MTkinomegafig}
\end{figure}

\begin{figure}[t]
\centering
\includegraphics[width=0.55\textwidth]{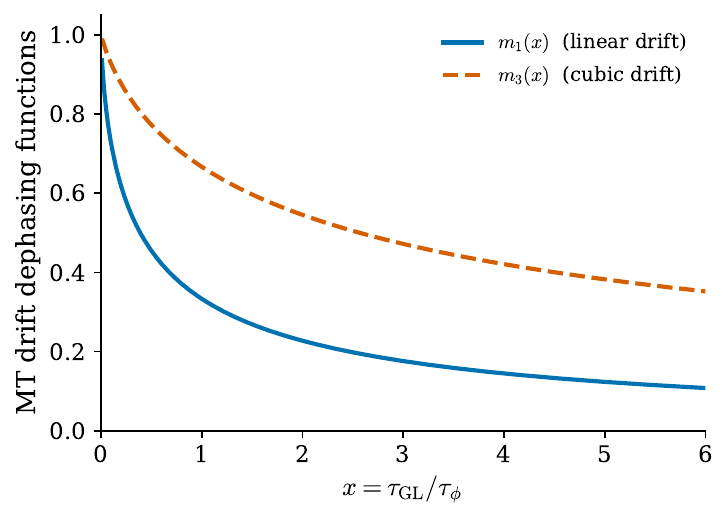}
\caption{Dephasing functions $m_1(x)$ and $m_3(x)$
[Eq.~\eqref{eq:m1m3}] of the kinetic (drift) MT channels.}
\label{fig:MTdriftfig}
\end{figure}

The finite-frequency generalization follows the same steps as the
thermodynamic sector, with the kinetic invariant entering through the
locked pair of insertions. Throughout this subsection we write the
kinetic invariant as
\begin{equation}
\rho(\q)=\mu\,R(\q),
\qquad
R_E\equiv\E\cdot\nabla R,
\label{eq:muR}
\end{equation}
where $\mu$ denotes the generic coupling constant: $\mu=u$ with
$R=\br\cdot\q$ for the linear drift, $\mu=u_3$ with $R=q^2(\br\cdot\q)$
for the cubic drift [Eq.~\eqref{eq:rho-C3v}], and the corresponding
strain couplings of Eq.~\eqref{eq:rho-D3h} for the $D_{3h}$ case; 
so that $\mu$ serves as the formal expansion parameter of the
linearization in the nonreciprocal perturbation. Along the driven trajectory
[$c(t)=\tfrac{2e}{\omega}\sin\omega t$; $R\equiv\rho/\mu$ the invariant
profile, $R_E\equiv\E\cdot\nabla R$],
the exponent of Eq.~\eqref{eq:genMT} acquires, besides the $\alpha$-drift
of Eq.~\eqref{eq:MTinner}, the rate terms
$+2\mu\alpha R\,(t_1{-}t_2)+2\mu(\alpha R_E+g_1R)\!\int_{t_2}^{t_1}c$, and the
noise vertex contributes $1-\mu R-\mu R_E\,c(t_2)$. Four
$\mathcal O(\mu E)$ sources result: the field part of the noise, the
static noise times the $\alpha$-drift, the field part of the rate
shift, and the static rate shift times the $\alpha$-drift (the cross
term of the exponential expansion). Their time integrals, the harmonic
projection, and the angular average give compact per-momentum kernels
[the Cooperon drift \eqref{eq:MTdrift} drops out of this sector, as
noted in Sec.~\ref{sec:MTkinetic}]. For the linear
drift,
\begin{equation}
j^{\rm dc}\propto
\frac{2z\big(4\alpha^2\beta+2\alpha\omega^2-\beta\omega^2\big)}
{\alpha(4\alpha^2+\omega^2)^2(4\beta^2+\omega^2)},
\qquad
j^{2\omega}\propto
\frac{z}{\alpha\,(2\alpha+i\omega)^2\,(2\beta+3i\omega)},
\label{eq:MTkin-kernels}
\end{equation}
and the cubic drift differs only by the radial weight $z\to z^2$; the
kinetic warping invariant vanishes at every frequency, since the
angular selection rule involves only the harmonic content of the
invariant. Defining the master functions [$\alpha=z+1$, $\beta=z+x$;
normalization matching the dc catalogue above]
\begin{align}
M_p^{\rm dc}(\nu,x)&=8\!\int_0^\infty\!\!dz\,
\frac{2z^{(p+1)/2}\big(4\alpha^2\beta+2\alpha\nu^2-\beta\nu^2\big)}
{\alpha(4\alpha^2+\nu^2)^2(4\beta^2+\nu^2)},
\nonumber\\
\mathbb M_p(\nu,x)&=8\!\int_0^\infty\!\!dz\,
\frac{z^{(p+1)/2}}{\alpha(2\alpha+i\nu)^2(2\beta+3i\nu)},
\label{eq:MTkin-M}
\end{align}
with $M_p^{c}=\re\,\mathbb M_p$, $M_p^{s}=-\im\,\mathbb M_p$
[$p=1$: linear drift; $p=3$: cubic drift], the finite-frequency
response reads
\begin{empheq}{align}
\delta\jj^{\rm PGE}_{\rm MT,kin}
&=\frac{e^3T\lambda_{\rm kin}}{2\pi}\,\tGL^{\,n_p}\;
M_p^{\rm dc}(\nu,x)\,\E(\E\cdot\br),
\label{eq:MTkinPGE}\\
\delta\jj^{\rm SHG}_{\rm MT,kin}
&=\frac{e^3T\lambda_{\rm kin}}{2\pi}\,\tGL^{\,n_p}\,
\big[M_p^{c}\cos2\omega t+M_p^{s}\sin2\omega t\big]\,
\E(\E\cdot\br),
\label{eq:MTkinSHG}
\end{empheq}
with $\lambda_{\rm kin}=u$, $n_1=2$ for the linear drift ($p{=}1$) and
$\lambda_{\rm kin}=u_3/D$, $n_3=1$ for the cubic drift ($p{=}3$), in
accordance with the dc catalogue of Sec.~\ref{sec:MTkinetic}, and the
usual strain substitutions for $D_{3h}$. The complex SHG amplitude has
an elementary closed form: writing the integrand of $\mathbb M_1$
(including the prefactor) as $z/[(z+1)(z+r_2)^2(z+r_3)]$ with
$r_2=1+i\nu/2$, $r_3=x+3i\nu/2$,
\begin{align}
\mathbb M_1(\nu,x)&=\frac{C}{r_2}-B\ln r_2-D\ln r_3,
\nonumber\\
B=\frac{r_3-r_2^{\,2}}{(1-r_2)^2(r_3-r_2)^2},
\qquad
C&=\frac{-r_2}{(1-r_2)(r_3-r_2)},
\qquad
D=\frac{-r_3}{(1-r_3)(r_2-r_3)^2}.
\label{eq:M1closed}
\end{align}
The low-frequency behavior of the kinetic sector shares the
qualitative features of the thermodynamic sector
(Fig.~\ref{fig:MTkinomegafig}). There is no $1/\nu$ pole
in the quadrature amplitude and no secular term at strict dc; here this
holds already at the level of the per-momentum kernels, because the
Cooperon drift does not enter the kinetic sector, and the quadrature
onset is linear,
\begin{equation}
M_1^{s}(\nu\to0,x)=\nu\,s_1(x),
\qquad
s_1(x)=\frac{2x^3-3x^2-24x\ln x+42x-18\ln x-41}{12(x-1)^4},
\qquad s_1(1)=\tfrac{5}{24}.
\label{eq:s1}
\end{equation}
The static limits split evenly,
\begin{equation}
M_p^{\rm dc}(0,x)=M_p^{c}(0,x)=\tfrac12\,m_p(x),
\label{eq:MTkin-statics}
\end{equation}
saturating the dc sum rule \eqref{eq:sumrule} in the quasistatic
($\cos^2$-rectification) pattern, as in Eq.~\eqref{eq:mdc0}. The
in-phase SHG amplitude $M_1^{c}$ changes sign at a frequency of order
$\tGL^{-1}$ that depends only weakly on the dephasing ratio
(Fig.~\ref{fig:MTkinomegafig}, middle panel), while the rectified
amplitude $M_1^{\rm dc}$ decays without changing sign. The kinetic and
thermodynamic MT channels are therefore distinguished not by the
presence or absence of these features but by their dephasing
dependence: the zero crossing of the kinetic in-phase second harmonic
moves only weakly with $x$, whereas the thermodynamic one,
$\nu_0(x)$ of Sec.~\ref{sec:MTresults}, varies severalfold over the
range of dephasing ratios shown in Fig.~\ref{fig:MTSHGfig}; the kinetic quadrature onset $s_1(x)$ is positive for
all $x$, whereas the thermodynamic onset $s_{\rm lin}(x)$ of
Eq.~\eqref{eq:slin} is negative for $x\lesssim0.04$; and the static
dephasing functions, $m_p(x)/2$ against $f(x)/8$, have different
profiles (Figs.~\ref{fig:MTdriftfig} and \ref{fig:MTPGEfig}). All results of
this subsection were verified by the same three-layer protocol,
including a thermodynamic-channel regression
and an expansion-free numerical evaluation of the master formula
(\ref{app:verification}).

\section{Circular polarization and the photovoltaic Hall effect}
\label{sec:PVHE}

The response to circularly and elliptically polarized light distinguishes
most clearly between the mechanisms assembled above. Take
\begin{equation}
\E(t)=\bm E_1\cos\omega t+\bm E_2\sin\omega t,
\qquad
\CE\equiv\bm E_1+i\bm E_2,
\qquad
\A(t)=-\frac{1}{\omega}\big(\bm E_1\sin\omega t-\bm E_2\cos\omega t\big),
\label{eq:elliptic}
\end{equation}
with circular polarization corresponding to
$\bm E_2=\pm\hat z\times\bm E_1$. The helicity-odd part of any
rectified response must be proportional to
$[\bm E_1\times\bm E_2]_z=\tfrac12\im[\mathcal E^*_x\mathcal E_y-
\mathcal E^*_y\mathcal E_x]\cdot$\,(sign convention), which reverses
under reversal of the light helicity.

\subsection{Helicity independence of the thermodynamic AL channel}
\label{sec:ALblind}

Repeating the AL expansion of Sec.~\ref{sec:ALtime} with the two
amplitudes of Eq.~\eqref{eq:elliptic}, the rectified part of the
Langevin factor evaluates to
\begin{equation}
Y^{\rm dc}_2=
\frac{12\alpha^2+\omega^2}{\alpha^3(4\alpha^2+\omega^2)^2}\,
\big|e\,\CE\cdot\nabla\alpha\big|^2
-\frac{e^2\,\re\big[\mathcal E^*_i\mathcal E_j\big]\,
\partial_i\partial_j\alpha}
{\alpha^2(4\alpha^2+\omega^2)} ,
\label{eq:ALelliptic}
\end{equation}
with no $\im[\mathcal E^*_i\mathcal E_j]$ term: the cross
coefficients of the two quadratures cancel identically (we verified
this both symbolically and by an expansion-free numerical evaluation of
the AL kernel). Equation \eqref{eq:ALelliptic} depends only on
$|\CE\cdot\nabla\alpha|^2$ and on the symmetric combination
$\re[\mathcal E^*_i\mathcal E_j]$: the relaxational
(real-coefficient) TDGL dynamics of the AL channel is helicity
independent at second order, and its circular photogalvanic response vanishes
for any pair spectrum $\delta\alpha$. This is a dynamical rather than
a symmetry statement. The helicity-odd invariant is allowed by
symmetry (Sec.~\ref{sec:symmetry}), but its coefficient vanishes because
the white-noise AL kernel has no phase memory, so the rectified response
depends only on the instantaneous intensity of the drive along the
trajectory.

Two clarifications are in order. First, the statement concerns the
helicity-odd part of the response only. Under circularly polarized light,
$\re[\mathcal E^*_i\mathcal E_j]=E^2\delta_{ij}$ and
$|\CE\cdot\nabla\alpha|^2=E^2|\nabla\alpha|^2$, with $E$ the amplitude
of each field component, so the rectified AL current equals the sum of
the responses to the two orthogonal linear polarizations. It is
independent of the helicity but it does not vanish whenever a polar
vector is available: for the $C_{3v}$ invariant it equals
$(e^3T\alpha_3/\pi D)\tGL^2\,\mathfrak g(\nu)\,4E^2\br$ [from
Eq.~\eqref{eq:C3vALPGE}, since $\bm F(\hat x)+\bm F(\hat y)=4E^2\br$],
and the strain channel of the $D_{3h}$ problem gives the same with
$\br\to\eps$. For the warping channel the trigonal doublet $\bm F_w$
averages to zero under circular polarization, as required by
symmetry: in the unstrained $D_{3h}$ crystal no in-plane vector can be
formed from $B_z$ together with either $E^2$ or the helicity
$[\bm E_1\times\bm E_2]_z$ (Sec.~\ref{sec:symmetry}). This accounts for
the observation of Ref.~\cite{Parafilo2022} that the
fluctuation PGE of the unstrained Ising superconductor disappears for
circular light. Second, the vanishing of the helicity-odd AL current is a
theorem only under the assumptions of the model: relaxational dynamics
with real Ginzburg-Landau coefficients, momentum-independent (white)
noise, no built-in dc field, and second order in the drive. Helicity-odd
fluctuation photocurrents can arise outside these assumptions. The ratchet
Hall effect of Ref.~\cite{Parafilo2025} requires an imaginary part of the
TDGL relaxation coefficient (particle-hole asymmetry) together with a
spatial modulation of $T_c$, and the photovoltaic Hall effect of
Ref.~\cite{BoevKovalev2024} is third order in the field and requires a
dc bias. These mechanisms are complementary to the present statement and
not in conflict with it. The AL
second harmonic, by contrast, couples to the rotating component of the
drive through the complex amplitude
\begin{equation}
Y^{2\omega}_2=
\frac{(3\alpha+i\omega)\,
\big(e\,\CE^*\!\cdot\nabla\alpha\big)^2}
{2\alpha^3(\alpha+i\omega)(2\alpha+i\omega)^2}
-\frac{e^2\,\mathcal E^*_i\mathcal E^*_j\,\partial_i\partial_j\alpha}
{4\alpha^2(\alpha+i\omega)(2\alpha+i\omega)},
\label{eq:AL2w}
\end{equation}
whose real projection reproduces
$\mathfrak g_c,\mathfrak g_s$ for linear polarization.

\subsection{Helicity-odd rectification in the MT channel}
\label{sec:MTcirc}

The MT channel behaves differently: its kernel retains phase memory
through the Cooperon time structure, and the rectified response
acquires a helicity-odd term. Carrying the elliptic drive through the
expansion of Sec.~\ref{sec:MTthermo}, the dc current becomes
\begin{align}
\jj^{\rm dc}_{\rm MT}=&16e^3TD\intq\Big\{c_{\rm dc}\;\re\big[\CE\,(\CE^*\!\cdot\nabla\alpha)\big]
-c_{\rm circ}\;
\im\big[\CE\,(\CE^*\!\cdot\nabla\alpha)\big]\nonumber \\ 
&+\delta\alpha_\q\Big(K_{\rm dc}\;\re\big[\CE\,(\CE^*\!\cdot\nabla\beta)\big]
-K_{\rm circ}\;\im\big[\CE\,(\CE^*\!\cdot\nabla\beta)\big]\Big)\Big\},
\nonumber\\
c_{\rm circ}&=\frac{4\alpha\beta+\omega^2}
{\omega\,\alpha(4\alpha^2{+}\omega^2)(4\beta^2{+}\omega^2)} ,
\qquad
K_{\rm circ}=-\frac{1}{4\alpha^2\omega(\beta^2+\omega^2)} ,
\label{eq:MTelliptic}
\end{align}
with $c_{\rm dc}$ and $K_{\rm dc}=1/4\alpha^2\beta(\beta^2+\omega^2)$ the
rectified kernels of Eq.~\eqref{eq:MTkernels}; the $K$ terms are the
Cooperon-drift contributions, which for linear polarization reduce to the
second bracket of Eq.~\eqref{eq:MTkernels}.
Using the vector identity
$-\im\big[\CE(\CE^*\!\cdot\nabla\alpha)\big]
=\nabla\alpha\times[\bm E_1\times\bm E_2]$ and the reduction
\eqref{eq:MTred},
\begin{equation}
\delta\jj^{\,\rm circ}_{\rm MT}
=\frac{4e^3T\lambda}{\pi D}\,\tGL^2\,
m_{\rm circ}(\nu,x)\,[\bm E_1\times\bm E_2]_z\;[\br\times\hat z],
\qquad
\begin{cases}
\lambda=\eta B_z,\ \ \br=\eps & (D_{3h}),\\[2pt]
\lambda=\alpha_3,\ \ \br\times\hat z=-\bm B & (C_{3v}),
\end{cases}
\label{eq:MTcircboxed}
\end{equation}
with $m_{\rm circ}$ defined in Eq.~\eqref{eq:mfuncs} and plotted in
Fig.~\ref{fig:MTPGEfig} (right). The $1/\nu$ poles of the two terms
of $m_{\rm circ}$ cancel after momentum integration
[Eq.~\eqref{eq:residues}], and the helicity-odd response vanishes
linearly at low frequency,
\begin{align}
m_{\rm circ}(\nu\to0,x)&=\nu\,s_{\rm circ}(x),\qquad s_{\rm circ}(1)=\tfrac1{240},
\nonumber\\
s_{\rm circ}(x)&=\frac{-x^4-26x^3+18x^2+10x-1+12x^2(x+2)\ln x}{48\,x\,(x-1)^5},
\label{eq:mcirc-res}
\end{align}
with $s_{\rm circ}\simeq1/48x$ for $x\ll1$ and a sign reversal of the
onset at $x\simeq5.4$. For weak pair breaking the slope is controlled by
$\tau_\phi$, $m_{\rm circ}\simeq\nu/48x=\omega\tau_\phi/48$, and
the function peaks at $\omega\sim\tau_\phi^{-1}$ when
$\tau_\phi\gg\tGL$. 

The geometry of Eq.~\eqref{eq:MTcircboxed} is notable. In the strained TMD the helicity-odd current flows
perpendicular to the strain axis [$(\eps\times\hat z)\perp\eps$]
and reverses with the light helicity: a fluctuation
photovoltaic Hall effect, in the spirit of the nonlinear Hall
response of normal conductors, but generated here by preformed Cooper
pairs and locked to the pair of symmetry-breaking fields
$(B_z,\eps)$. In the Rashba geometry the current flows along the
in-plane magnetic field. The effect is second order in the drive,
whereas the photovoltaic Hall effect of Ref.~\cite{BoevKovalev2024} is
third order and requires a built-in dc field to supply the extra
vector. In the thermodynamic sector it is carried by the MT channel
alone, since the AL channel with white noise does not produce it
(Sec.~\ref{sec:ALblind}).

\subsection{Circular AL response from the nonreciprocal noise}
\label{sec:ALcirc}

The kinetic Lifshitz invariant changes the AL result qualitatively.
The FDT-locked noise gives the AL kernel the phase memory that it lacks
with white noise: repeating the elliptic-drive expansion of
the generalized formula \eqref{eq:genAL} with the linear drift
$\rho=u\,(\br\cdot\q)$, a helicity-odd rectified current appears,
\begin{equation}
\delta\jj^{\rm circ}_{\rm AL}
=-\frac{8e^3T\,u}{\pi}\,\tGL^{2}\;C_{\rm circ}(\nu)\,
[\bm E_1\times\bm E_2]_z\,[\br\times\hat{\bm z}],
\qquad
C_{\rm circ}(\nu)=\nu\!\int_0^\infty\!\!
\frac{\bar z\,d\bar z}{(\bar z{+}1)\,[4(\bar z{+}1)^2+\nu^2]^2},
\label{eq:ALcircboxed}
\end{equation}
with the master function shown in Fig.~\ref{fig:ALdrift} (right):
$C_{\rm circ}\simeq\nu/192$ at small $\nu$, a maximum at
$\nu\simeq2.2$, and the large-frequency tail
$C_{\rm circ}\simeq\pi/8\nu^2$, which follows from
$\nu\int_1^\infty dy\,(4y^2+\nu^2)^{-2}=\pi/8\nu^2+\mathcal O(\nu^{-3})$
and is approached slowly, with corrections of relative order
$\ln\nu/\nu$. The response carries the full
$\tGL^2$ singularity. For the Rashba geometry
$\br\times\hat z=-\bm B$: circularly polarized light drives a dc
current along the in-plane field, reversing with helicity.
Since the white-noise AL circular response vanishes identically for
arbitrary $\delta\alpha$, Eq.~\eqref{eq:ALcircboxed} identifies the
FDT-locked nonreciprocal noise as the leading AL mechanism of
circular photogalvanics near $T_c$ and, conversely, identifies the
circular PGE as a signature of the kinetic Lifshitz invariant that
is free of a thermodynamic fluctuation background: within the
$\tGL$-enhanced fluctuation response, and in the AL-dominated regime of
strong dephasing where the MT channel is suppressed, any helicity-odd
rectified signal measures the nonreciprocity of the pair relaxation. The
qualification matters in practice. Helicity-dependent photocurrents of
the normal-state band structure (the circular photogalvanic effect of
noncentrosymmetric conductors) and extrinsic contributions coexist in
real samples; they are distinguished from the fluctuation signal by the
absence of the critical temperature dependence on the scale $\tGL$ and
by the frequency profile $C_{\rm circ}(\omega\tGL)$, which is tied to
$T-T_c$. The same two criteria separate the fluctuation signal from the
helicity-dependent responses of other origin discussed below
Eq.~\eqref{eq:AL2w}.

Taken together, the two helicity-odd responses probe the two
sectors separately. The MT circular current
\eqref{eq:MTcircboxed} measures the thermodynamic invariant through the
quantum-interference channel, and the AL circular current
\eqref{eq:ALcircboxed} measures the kinetic invariant through the
paraconductivity channel. Both vanish linearly at low frequency, with
slopes $s_{\rm circ}(x)$ and $1/192$, respectively; the MT slope and peak
position depend on the dephasing ratio, with the peak at
$\omega\sim\tau_\phi^{-1}$ for weak pair breaking, whereas the AL profile
is dephasing independent and peaks at $\omega\simeq2.2\,\tGL^{-1}$. The
two can therefore be separated in a single experiment by frequency and
temperature sweeps.

\section{Symmetry validation of the vector structures}
\label{sec:symmetry}

Every vector structure obtained above can be checked against, and is
exhausted by, the point-group analysis of the photogalvanic tensor
$j_i=\chi_{ijk}E_jE_k$ (symmetrized in $jk$ for linear polarization),
with $\chi$ odd under inversion and, in our problems, linear in the
time-reversal-breaking field.

\paragraph{$C_{3v}$ (Rashba)} With an in-plane $\bm B$, the unique
polar in-plane vector linear in $B$ is $\br=\bm B\times\hat z$. The most
general rectified response built from one $\br$ and two $\E$'s that
respects the (effectively $C_{\infty v}$) symmetry is
\begin{equation}
\jj=\gamma_1\,\E(\E\cdot\br)+\gamma_2\,\br\,E^2 .
\label{eq:C3vgeneral}
\end{equation}
The computed channels realize this space with characteristic weight
ratios: the thermodynamic AL response has
$(\gamma_1{:}\gamma_2)=(2{:}1)$ [the $\bm F$ structure], the kinetic
(drift) AL response has $(4{:}{-}1)$, and the MT response is a pure
$\gamma_1$. Under the vertical mirror $x\to-x$ [$B_x\to B_x$,
$B_y\to-B_y$, i.e., $b_x\to-b_x$, $b_y\to b_y$] both structures
transform as proper polar vectors, as required. For elliptic
polarization exactly one additional helicity-odd invariant exists,
\begin{equation}
[\bm E_1\times\bm E_2]_z\,[\br\times\hat z]
=-[\bm E_1\times\bm E_2]_z\,\bm B ,
\label{eq:C3vcirc}
\end{equation}
realized by the MT circular term \eqref{eq:MTcircboxed} and by the
noise-induced AL term \eqref{eq:ALcircboxed}, with the white-noise AL
coefficient vanishing dynamically. The helicity-even part of the
response to circular light is the polarization average of
Eq.~\eqref{eq:C3vgeneral}, $(\gamma_1+2\gamma_2)E^2\br$, and is nonzero in
all channels [Sec.~\ref{sec:ALblind}].

\paragraph{$D_{3h}$ (Ising/TMD, field $B_z$)} Under $D_{3h}$ the
current doublet $(j_x,j_y)$ belongs to $E'$, the field $B_z$ to $A_2'$,
and the quadratic forms of the drive split as $E^2\in A_1'$ and
$(E_x^2-E_y^2,\,-2E_xE_y)\in E'$. Since $A_2'\otimes E'=E'$, exactly one
invariant linear in $B_z$ exists in the unstrained crystal:
\begin{equation}
\jj\propto B_z\,\bm F_w(\E),
\label{eq:D3hgeneral}
\end{equation}
fixing the warping channel completely, one coupling constant, as
found in Eq.~\eqref{eq:D3hALPGE}. Explicit checks: under $C_3$
[$\E\to R_{2\pi/3}\E$] the doublet transforms as
$\bm F_w\to R_{-4\pi/3}\bm F_w=R_{2\pi/3}\bm F_w$, so the current
rotates as a vector; under $x\to-x$, $B_z\to-B_z$ and
$(F_{w,x},F_{w,y})\to(F_{w,x},-F_{w,y})$, giving
$(j_x,j_y)\to(-j_x,j_y)$ as required. With the strain doublet
$\eps\in E'$ two further invariants appear at
$\mathcal O(B_z\varepsilon)$, namely $\E(\E\cdot\eps)$ and
$\eps\,E^2$ [both present in the AL response as $\bm F_\eps$, and with
different weights in the kinetic sector; only the first in MT], plus
the helicity-odd combination
$[\bm E_1\times\bm E_2]_z\,[\eps\times\hat z]$ realized by the MT
circular response. Because the kinetic invariants $\rho$ carry
the same transformation properties as their thermodynamic
counterparts $\delta\alpha$ [Eqs.~\eqref{eq:onsager}--\eqref{eq:Gamma}],
they cannot enlarge this list: all computed structures (AL and MT,
thermodynamic and kinetic, linear and circular) exhaust
the symmetry-allowed set. The
dynamics determines the values and frequency dependences of the
coefficients that symmetry leaves free: the weight ratios
$(2{:}1)$ vs $(4{:}{-}1)$ vs pure-$\gamma_1$, the vanishing of the
white-noise AL circular coefficient, and the $\tGL$-power hierarchy
between the cubic and the linear invariants.

\section{Summary and outlook}
\label{sec:summary}

We have constructed a complete theory of the second-order optical
response of two-dimensional noncentrosymmetric superconductors in the
fluctuation regime, unifying the photogalvanic effect, second-harmonic
generation, and the photovoltaic Hall effect within a single
TDGL-based framework extended in two directions: to the
quantum-interference (Maki-Thompson) channel alongside the
paraconductivity (Aslamazov-Larkin) one, and to kinetic Lifshitz
invariants, the Onsager-allowed, FDT-locked nonreciprocity of the
pair relaxation, alongside the thermodynamic invariants of
the pair spectrum. The main results can be summarized as follows.

(i) \emph{Master formulas and structure.} The generalized Schmid
formula \eqref{eq:genAL} and the MT master formula \eqref{eq:genMT}
hold at arbitrary drive frequency below the fermionic scale
$\omega\sim T$ that limits the relaxational TDGL description, and to
linear order in the nonreciprocal perturbations. Two exact properties
organize the theory: the equilibrium state is strictly Gibbsian, the
kinetic invariant drops out of all static quantities by the FDT lock;
and the odd-in-field linear response vanishes in both channels, so all
nonreciprocity is intrinsically nonlinear. In the MT channel the
field dependence of the electronic Cooperon pole enters at the same order
as that of the pair propagator and is retained throughout; it removes the
$1/\omega$ poles of the quadrature and helicity-odd amplitudes from the
momentum-integrated response and enforces the even static split between
the rectified and the in-phase second-harmonic amplitudes.

(ii) \emph{Thermodynamic sector.} The AL response of any cubic
Lifshitz invariant is governed by the universal dimensionless functions
$\mathfrak g,\mathfrak g_c,\mathfrak g_s$ of
Eqs.~\eqref{eq:gAL}--\eqref{eq:gs}, with the $\tGL^2$ singularity and
identical frequency profiles for the $C_{3v}$ invariant and for both
$D_{3h}$ channels (warping and strain); the profile agrees with
the independent calculation of Ref.~\cite{Parafilo2022} in the corresponding channel. The MT
response introduces the dephasing ratio $x=\tGL/\tau_\phi$ through the
master functions \eqref{eq:mfuncs}, carries the same $\tGL^2$
singularity as the AL response with the static dephasing function
$f(x)$ of Eq.~\eqref{eq:mdc0} [comparable to AL, without the logarithmic
enhancement of the linear MT conductivity], has an in-phase second
harmonic that changes sign at a dephasing-dependent frequency, vanishes
identically for the warping invariant (angular selection rule), and
realizes a pure $\E(\E\cdot\br)$ polarization structure.

(iii) \emph{Kinetic sector.} Kinetic invariants with the same
momentum structure as a thermodynamic invariant produce the same vector
structures suppressed by one power of $\tGL$; the linear kinetic drift,
which has no non-removable thermodynamic counterpart, contributes
to both channels at the full $\tGL^2$ order: in the AL channel with
distinctive polarization weights $(4{:}{-}1)$ and its own frequency
profile, with the anisotropy ratio locked near its dc value $-\tfrac14$
--- opposite in sign to the thermodynamic $(2{:}1)$ structure; in the
MT channel with the dephasing function $m_1(x)$ and the pure
$\E(\E\cdot\br)$ structure shared with the thermodynamic term. In the MT
channel the kinetic master functions are obtained in closed form
[Eqs.~\eqref{eq:MTkin-M}--\eqref{eq:M1closed}]; the kinetic and
thermodynamic MT responses are distinguished by their dephasing
functions and by the dephasing dependence of the zero crossing of the
in-phase second harmonic, which is weak in the kinetic and strong in the
thermodynamic channel.

(iv) \emph{Circular responses.} With white noise the AL channel is
helicity independent, its circular PGE vanishes for any pair spectrum,
while the MT channel supports a helicity-odd rectified current
\eqref{eq:MTcircboxed} flowing transverse to the strain axis
($D_{3h}$) or along the field ($C_{3v}$): a fluctuation photovoltaic
Hall effect at second order in the drive, which vanishes linearly at
low frequency and peaks at $\omega\sim\tau_\phi^{-1}$ for weak pair
breaking. The kinetic invariant restores the helicity dependence of
the AL channel [Eq.~\eqref{eq:ALcircboxed}], making the circular PGE
a signature of nonreciprocal pair kinetics that is free of a
thermodynamic fluctuation background in the AL-dominated regime.

(v) \emph{Scope and expected magnitudes.} The results above are exact
consequences of the relaxational TDGL model with FDT-locked
momentum-dependent kinetic coefficient, within the Gaussian fluctuation
regime and for $\omega\ll T$; the symmetry classification of the
invariants, the structure of the master formulas, the vector structures,
the $\tGL$ hierarchy, and the dimensionless master functions do not
depend on microscopic input. What does require microscopic input are the
values of the coupling constants $\alpha_3,\kappa,\eta$ and
$u,u_3,\tilde\kappa,\tilde\eta,\upsilon$, including their disorder
dependence, and the pair-breaking time $\tau_\phi$; the statements about
the relative size of the thermodynamic and kinetic sectors in a given
material therefore remain predictions conditional on these constants.
The magnitude of the effect can be expressed through a single
nonreciprocity length. Writing $\ell\equiv\alpha_3|\bm B|/D$ for the
$C_{3v}$ invariant [$\delta\alpha/Dq^2=\ell\,q\cos\theta_{\q\br}$, so that
$\ell/\xi_{\rm GL}$ is the relative asymmetry of the pair spectrum at the
fluctuation momentum], and restoring $\hbar$ and $k_B$, the rectified AL
current \eqref{eq:C3vALPGE} reads
\begin{equation}
\delta j^{\rm PGE}_{\rm AL}=\frac{e^3k_BT\,\ell\,\tGL^2E^2}{\pi\hbar^3}\,
\mathfrak g(\omega\tGL)\,\hat F,
\qquad
\tGL=\frac{\pi\hbar}{8k_B(T-T_c)},
\label{eq:estimate}
\end{equation}
with $\hat F=3$ for $\E\parallel\br$. For $T_c=10\,$K and
$\epsilon=0.1$ ($\tGL\simeq3\,$ps, $\xi_{\rm GL}\simeq30\,$nm) the
weak-field condition of Sec.~\ref{sec:genMT} reads
$E\ll\hbar/e\xi_{\rm GL}\tGL\simeq70\,$V/cm. Taking $\ell=0.1\,$nm (a
relative asymmetry $\ell/\xi_{\rm GL}\sim3\times10^{-3}$),
$E=10\,$V/cm, and $\omega\tGL\ll1$ gives
$\delta j\simeq2\times10^{-5}\,$A/m, i.e., about $2\,$nA for a sample
$100\,\mu$m wide, growing as $(T-T_c)^{-2}$ while $\omega\tGL\ll1$. A
measure independent of the sample is the ratio to the linear
paraconductivity current \eqref{eq:lin},
\begin{equation}
\frac{\delta j^{\rm PGE}_{\rm AL}}{j^{\rm lin}_{\rm AL}}
=2\,\mathfrak g(\omega\tGL)\,\hat F\;\frac{\ell}{\xi_{\rm GL}}\;
\frac{eE\,\xi_{\rm GL}\tGL}{\hbar},
\label{eq:ratio}
\end{equation}
the product of the spectral asymmetry and the dimensionless field. At
$\omega\tGL\ll1$ and $\hat F=3$ the numerical prefactor is $\tfrac14$, and
the ratio is about $10^{-4}$ for the parameters above. The condition
$\omega\tGL\ll1$ is $\epsilon\gg\pi\hbar\omega/8k_BT_c$, which for
$T_c=10\,$K reads $\epsilon\gg0.02$ at $10\,$GHz and $\epsilon\gg0.2$ at
$100\,$GHz; the latter frequency is already close to the limit
$\hbar\omega\ll k_BT$ of the TDGL description. For $\omega\tGL\gg1$ the
rectified response is frequency limited,
$\tGL^2\mathfrak g\to\omega^{-2}$, and independent of $T-T_c$. The critical
enhancement is therefore accessible in the microwave and sub-THz range
for $T_c\sim10\,$K, and the frequency profiles of
Figs.~\ref{fig:ALfig}--\ref{fig:MTkinomegafig} describe the crossover
between the two regimes. The MT contributions are of the same order with
$\mathfrak g\to4m_{\rm dc}$, and the helicity-odd currents are smaller by
the factors $C_{\rm circ}$ and $m_{\rm circ}$ of
Figs.~\ref{fig:ALdrift} and \ref{fig:MTPGEfig}.

Several extensions suggest themselves naturally. A microscopic
derivation of the kinetic coefficients $u$, $u_3$, $\tilde\kappa$,
$\tilde\eta$ from the frequency dependence of the pair susceptibility
of specific band models, including their disorder renormalization,
which for the thermodynamic warping invariant is known to be strong,
would convert the present phenomenology into quantitative material
predictions for MoS$_2$-class and Rashba systems \cite{TdM2026a,Liu2026}. The
density-of-states channel of the fluctuation response remains
unaddressed at finite frequency in the present context; for the
nonlinear Hall response of fluctuating pairs it was found to be
subdominant to the AL term \cite{Dong2025}. Embedding a
built-in dc field into the master formulas [$\E\to\bm F+\E(t)$] would
connect our second-order photovoltaic Hall response to the third-order
effect of Ref.~\cite{BoevKovalev2024} within one formalism, including
its temperature dependence near $T_c$ and dephasing corrections.
The behavior of these responses below $T_c$ is, to our knowledge,
not established. Once a condensate coexists with the fluctuations, the
AL and MT channels must evolve into condensate and collective-mode
responses, including the nonreciprocal superfluid weight and the Higgs
and phase modes with their finite-momentum electrodynamics
\cite{WatanabeDaidoYanase2022,SunMillis2020}. How the PGE, SHG, and
photovoltaic Hall currents computed here change across the transition,
which divergences are cut by the condensate, and what replaces the
dephasing regularization of the MT channel are open problems that we
leave for future work. A related but distinct effect in the
superconducting state is the inverse Faraday effect, by which circularly
polarized light transfers angular momentum to the condensate and can
switch between current-carrying states \cite{Croitoru2023}. That
mechanism acts on an existing condensate, produces a magnetization or a
circulating current rather than a dc transport current, and does not
require broken inversion symmetry; the helicity-odd fluctuation
photocurrents of Sec.~\ref{sec:PVHE} exist above $T_c$, are dc transport
currents second order in the field, and are odd in the symmetry-breaking
fields ($\bm B$, or $B_z$ and $\eps$) that the Lifshitz invariants
require. Whether the fluctuation helicity-odd currents connect
continuously to an inverse-Faraday-type response of the nonreciprocal
condensate below $T_c$ is part of the open problem stated above. The
finite-momentum generalization relevant to near-field THz probes
\cite{vonHoegen2025} is a further natural direction. On the
experimental side, the predictions are concrete: characteristic
frequency profiles on the scale $\omega\sim\tGL^{-1}$ with
$(T-T_c)^{-2}$ enhancement in the regime $\omega\tGL\ll1$,
polarization weights that identify the channel and sector, a
dephasing-sensitive MT component, and helicity-odd currents with
prescribed geometry, all accessible to microwave and THz
photocurrent and harmonic-generation spectroscopy of gated MoS$_2$,
strained TMD films, and Rashba heterostructures near their
superconducting transitions.
\section*{Acknowledgments}

The work of J. T. M. was supported in part by the Coordenação de Aperfeiçoamento de Pessoal de Nível Superior - Brasil (CAPES) - Finance Code 001 and by the NSF Quantum Leap Challenge Institute for Hybrid Quantum Architectures and Networks Grant No. OMA-2016136. The work of A. L. was supported by NSF Grant No. DMR-2452658 and H. I. Romnes Faculty Fellowship provided by the University of Wisconsin-Madison Office of the Vice Chancellor for Research and Graduate Education with funding from the Wisconsin Alumni Research Foundation. The authors acknowledge the use of Claude (Anthropic, model Fable 5.2) with manuscript preparation, which includes in particular numerics, graphics, and symbolic verification of analytics. All results were conceptualized, checked, and validated by the authors.

\appendix

\section{Angular reduction: derivation}
\label{app:reduction}

All angular averages follow from the 2D moments
\begin{equation}
\langle q_iq_j\rangle=\frac{q^2}{2}\,\delta_{ij},
\qquad
\langle q_iq_jq_kq_l\rangle
=\frac{q^4}{8}\big(\delta_{ij}\delta_{kl}
+\delta_{ik}\delta_{jl}+\delta_{il}\delta_{jk}\big),
\label{eq:moments}
\end{equation}
and the corresponding sixth-order moment with weight $q^6/48$ and
fifteen pairings; $\langle\cdots\rangle$ denotes the average over the
orientation of $\q$ at fixed $q$.

\paragraph{Case (a): $\delta\alpha=\lambda q^2(\br\cdot\q)$, AL
integrand} The $\mathcal O(\lambda)$ part of the AL integrand
[Eqs.~\eqref{eq:ALPGE}--\eqref{eq:ALSHG}] receives four insertions:
$\delta\alpha$ in the argument of $C_{A,B}(\alpha)$ [chain rule,
producing $C'_{A,B}\,\delta\alpha$], in the two gradient factors
$(\E\cdot\nabla\alpha)^2\to2(\E\cdot2D\q)(\E\cdot\nabla\delta\alpha)$,
in the Hessian $E_iE_j\partial_i\partial_j\delta\alpha$, and in the
current vertex $\partial_i\delta\alpha$. Using
$\nabla\delta\alpha=\lambda[2\q(\br\cdot\q)+q^2\br]$,
$E_iE_j\partial_i\partial_j\delta\alpha
=\lambda[2(\br\cdot\q)E^2+4(\E\cdot\q)(\E\cdot\br)]$, and evaluating each contraction with the moments
\eqref{eq:moments}, the angular average assembles into the two radial
weights quoted in Eq.~\eqref{eq:PS}:
\begin{equation}
\big\langle\cdots\big\rangle_i
=P(z)\,(\E\cdot\br)E_i+S(z)\,E^2 b_i,
\qquad
\begin{aligned}
P&=\big[2C_A'z^3+10C_Az^2+4C_Bz\big]/D^2,\\
S&=\big[C_A'z^3+5C_Az^2+2C_B'z^2+6C_Bz\big]/D^2 .
\end{aligned}
\end{equation}
Both weights were verified against high-precision numerical angular
integration (ten digits). Because all boundary terms vanish
[$C_{A,B}$ decay at least as $z^{-3}$ at large $z$ and are regular at
$z=0$ for $\Delta>0$], integration by parts in $z$ is legitimate and
yields $\int_0^\infty P\,dz=2\int_0^\infty S\,dz$, whence the fixed
vector structure $\bm F=2\E(\E\cdot\br)+\br E^2$ of
Eq.~\eqref{eq:F} after radial integration only.

\paragraph{Case (b): warping $\delta\alpha=\lambda q_x(q_x^2-3q_y^2)$}
The same procedure with the third-harmonic form factor
[$q_x(q_x^2-3q_y^2)=q^3\cos3\theta$] gives
\begin{equation}
\big\langle\cdots\big\rangle
=\frac{6z\,[C_A z+C_B]}{D^2}\,\bm F_w(\E),
\end{equation}
i.e., three times the radial weight \eqref{eq:Sint} with the trigonal
doublet $\bm F_w=(E_x^2-E_y^2,-2E_xE_y)$; no integration by parts is
needed here, the $C_3$-covariant structure emerges pointwise in $z$.

\paragraph{Case (c): MT integrand} The MT current has a single
gradient factor, $\intq C_M(\alpha,\beta)\E(\E\cdot\nabla\alpha)\cdot
(\text{vertex }\E)$, hence only two $\mathcal O(\lambda)$ insertions:
the chain rule in $C_M$ [at fixed $\beta$, the Cooperon pole carries
no Lifshitz invariant] and the gradient/vertex factors. For the
$C_{3v}$-type invariant the moments \eqref{eq:moments} give
Eq.~\eqref{eq:MTred},
$\mathcal R[C]=(\partial C/\partial\alpha)_\beta z^2+2Cz$. The
Cooperon-drift term of Eq.~\eqref{eq:MTkernels} has the structure
$K(\alpha,\beta)\,(\E\cdot\nabla\beta)\,\delta\alpha\cdot(\text{vertex }\E)$
with $\nabla\beta=2D\q$; its only $\mathcal O(\lambda)$ insertion is
$\delta\alpha$ itself, and the same first moment gives
$\langle(\E\cdot\nabla\beta)\,\delta\alpha\rangle=\lambda z^2e(\E\cdot\br)$
(units $D=1$), i.e., the radial weight $z^2K$ added to $\mathcal R[C]$ in
Eq.~\eqref{eq:mfuncs}. For the
warping invariant, the required averages
$\langle\delta\alpha\,q_i\rangle$ and
$\langle\partial_i\delta\alpha\rangle$ vanish because
$\cos3\theta$ has no overlap with the first harmonic. The same moments applied to the elliptic
drive produce the identity
$-\im[\CE(\CE^*\!\cdot\nabla\alpha)]
=\nabla\alpha\times[\bm E_1\times\bm E_2]$ used in
Sec.~\ref{sec:MTcirc}.

\section{Verification methodology}
\label{app:verification}

Given the length of the calculations, all results in this paper were
established with a three-layer verification protocol, scripted and
reproducible. (1)~\emph{Symbolic}: every time integral, Fourier
projection, angular average, and static limit was re-derived by
computer algebra; angular averages via exact trigonometric moments;
radial integrals in closed form where available. (2)~\emph{Numeric
closed-form checks}: every closed-form expression was evaluated against
high-precision quadrature of its defining integral at multiple
parameter points, with residuals at machine precision.
(3)~\emph{End-to-end}: the master formulas
\eqref{eq:genAL} and \eqref{eq:genMT} were evaluated numerically
without any expansion in the field --- exact time-ordered
exponents on quadrature grids at small $E$, with Fourier extraction of
the dc and $2\omega$ components and even/odd separation in $\pm\E$ ---
and the extracted harmonics reproduce the analytic kernels.

The linear-response no-go was confirmed in both channels, the dc sum rule
\eqref{eq:sumrule} holds term by term for every channel, and the
helicity independence of the white-noise AL channel was confirmed
end-to-end with an exact quadratic trial spectrum. Independent
cross-checks against the literature: the AL frequency profile agrees
with Ref.~\cite{Parafilo2022} (Fig.~\ref{fig:ALfig}), and the
static limits of the AL channel reproduce the dc magnetochiral results of
Refs.~\cite{TdM2026a,TdM2026b}.

\bibliographystyle{elsarticle-num}
\bibliography{refs}

\begin{thebibliography}{10}
\expandafter\ifx\csname url\endcsname\relax
  \def\url#1{\texttt{#1}}\fi
\expandafter\ifx\csname urlprefix\endcsname\relax\def\urlprefix{URL }\fi
\expandafter\ifx\csname href\endcsname\relax
  \def\href#1#2{#2} \def\path#1{#1}\fi

\bibitem{Basov2017}
D.~N. Basov, R.~D. Averitt, D.~Hsieh, Towards properties on demand in quantum
  materials, Nature Materials 16 (2017) 1077.

\bibitem{Torre2021}
A.~de~la Torre, D.~M. Kennes, M.~Claassen, S.~Gerber, J.~W. McIver, M.~A.
  Sentef, Colloquium: Nonthermal pathways to ultrafast control in quantum
  materials, Reviews of Modern Physics 93 (2021) 041002.

\bibitem{Orenstein2021}
J.~Orenstein, J.~E. Moore, T.~Morimoto, D.~H. Torchinsky, J.~W. Harter,
  D.~Hsieh, Topology and symmetry of quantum materials via nonlinear optical
  responses, Annual Review of Condensed Matter Physics 12 (2021) 247.

\bibitem{Nakamura2020}
S.~Nakamura, H.~Iida, Y.~Murotani, R.~Matsunaga, H.~Terai, R.~Shimano, Infrared
  activation of the {Higgs} mode by supercurrent injection in superconducting
  nbn, Physical Review Letters 125 (2020) 097004, nonreciprocal terahertz
  second-harmonic generation under supercurrent injection.

\bibitem{Matsunaga2014}
R.~Matsunaga, N.~Tsuji, H.~Fujita, A.~Sugioka, K.~Makise, Y.~Uzawa, H.~Terai,
  Z.~Wang, H.~Aoki, R.~Shimano, Light-induced collective pseudospin precession
  resonating with {Higgs} mode in a superconductor, Science 345 (2014) 1145.

\bibitem{ShimanoTsuji2020}
R.~Shimano, N.~Tsuji, Higgs mode in superconductors, Annual Review of Condensed
  Matter Physics 11 (2020) 103.

\bibitem{Katsumi2018}
K.~Katsumi, N.~Tsuji, Y.~I. Hamada, R.~Matsunaga, J.~Schneeloch, R.~D. Zhong,
  G.~D. Gu, H.~Aoki, Y.~Gallais, R.~Shimano, Higgs mode in the $d$-wave
  superconductor {Bi$_2$Sr$_2$CaCu$_2$O$_{8+x}$} driven by an intense terahertz
  pulse, Physical Review Letters 120 (2018) 117001.

\bibitem{Nakamura2024}
S.~Nakamura, H.~Matsumoto, H.~Ogawa, T.~Kobayashi, F.~Nabeshima, A.~Maeda,
  R.~Shimano, Picosecond trajectory of two-dimensional vortex motion in
  {${\mathrm{FeSe}}_{0.5}{\mathrm{Te}}_{0.5}$} visualized by terahertz second
  harmonic generation, Phys. Rev. Lett. 133 (2024) 036004.

\bibitem{Sekiguchi2024}
F.~Sekiguchi, H.~Narita, H.~Hirori, T.~Ono, Y.~Kanemitsu, Anomalous behavior of
  critical current in a superconducting film triggered by {DC} plus terahertz
  current, Nature Communications 15~(1) (2024) 4435.

\bibitem{Hashimoto2025}
M.~Hashimoto, T.~Konoike, T.~Kobayashi, S.~Hoshino, T.~Kawada, T.~Yokouchi,
  S.~Uji, A.~Maeda, Y.~Shiomi, Nonlinear planar {Hall} effect from
  superconducting vortex motion, Phys. Rev. Lett. 135 (2025) 136001.

\bibitem{Fausti2011}
D.~Fausti, R.~I. Tobey, N.~Dean, S.~Kaiser, A.~Dienst, M.~C. Hoffmann, S.~Pyon,
  T.~Takayama, H.~Takagi, A.~Cavalleri, Light-induced superconductivity in a
  stripe-ordered cuprate, Science 331 (2011) 189.

\bibitem{Mitrano2016}
M.~Mitrano, A.~Cantaluppi, D.~Nicoletti, S.~Kaiser, A.~Perucchi, S.~Lupi, P.~D.
  Pietro, D.~Pontiroli, M.~Ricc{\`o}, S.~R. Clark, D.~Jaksch, A.~Cavalleri,
  Possible light-induced superconductivity in {K$_3$C$_{60}$} at high
  temperature, Nature 530 (2016) 461.

\bibitem{Cavalleri2018}
A.~Cavalleri, Photo-induced superconductivity, Contemporary Physics 59 (2018)
  31.

\bibitem{Bilbro2011}
L.~S. Bilbro, R.~V. Aguilar, G.~Logvenov, O.~Pelleg, I.~Bo{\v z}ovi{\'c}, N.~P.
  Armitage, Temporal correlations of superconductivity above the transition
  temperature in {La$_{2-x}$Sr$_x$CuO$_4$} probed by terahertz spectroscopy,
  Nature Physics 7 (2011) 298.

\bibitem{Zhou2026}
H.~Zhou, Z.~Chen, A.~A. Varlamov, A.~Glatz, Y.~Yerin, Microwave signature of
  the emerging {Abrikosov} lattice above {$H_{c2}$}, Physical Review Letters
  137 (2026) 036001.

\bibitem{vonHoegen2025}
A.~von Hoegen, T.~Tai, C.~J. Allington, M.~Yeung, J.~Pettine, M.~H. Michael,
  E.~V. Bostr{\"o}m, X.~Cui, K.~Torres, A.~E. Kossak, B.~Lee, G.~S.~D. Beach,
  G.~Gu, A.~Rubio, P.~Kim, N.~Gedik, Visualizing a terahertz superfluid plasmon
  in a two-dimensional superconductor (2025).
\newblock \href {http://arxiv.org/abs/2506.08204} {\path{arXiv:2506.08204}}.

\bibitem{SunMillis2020}
Z.~Sun, M.~M. Fogler, D.~N. Basov, A.~J. Millis, Collective modes and terahertz
  near-field response of superconductors, Physical Review Research 2 (2020)
  023413.

\bibitem{XMM2019}
T.~Xu, T.~Morimoto, J.~E. Moore, Nonlinear optical effects in
  inversion-symmetry-breaking superconductors, Physical Review B 100 (2019)
  220501.

\bibitem{WatanabeDaidoYanase2022}
H.~Watanabe, A.~Daido, Y.~Yanase, Nonreciprocal optical response in
  parity-breaking superconductors, Physical Review B 105 (2022) 024308.

\bibitem{TanakaWatanabeYanase2023}
H.~Tanaka, H.~Watanabe, Y.~Yanase, Nonlinear optical responses in
  noncentrosymmetric superconductors, Physical Review B 107 (2023) 024513.

\bibitem{TanakaWatanabeYanase2024}
H.~Tanaka, H.~Watanabe, Y.~Yanase, Nonlinear optical responses in
  superconductors under magnetic fields: Quantum geometry and topological
  superconductivity, Physical Review B 110 (2024) 014520.

\bibitem{Raj2024}
A.~Raj, S.~Kaushik, et~al., Nonlinear optical responses in multiorbital
  topological superconductors, Physical Review B 109 (2024) 184514.

\bibitem{Wakatsuki2017}
R.~Wakatsuki, Y.~Saito, S.~Hoshino, Y.~M. Itahashi, T.~Ideue, M.~Ezawa,
  Y.~Iwasa, N.~Nagaosa, Nonreciprocal charge transport in noncentrosymmetric
  superconductors, Science Advances 3 (2017) e1602390.

\bibitem{Itahashi2020}
Y.~M. Itahashi, T.~Ideue, Y.~Saito, S.~Shimizu, T.~Ouchi, T.~Nojima, Y.~Iwasa,
  Nonreciprocal transport in gate-induced polar superconductor {SrTiO$_3$},
  Science Advances 6 (2020) eaay9120.

\bibitem{Edelstein1996}
V.~M. Edelstein, The {Ginzburg-Landau} equation for superconductors of polar
  symmetry, Journal of Physics: Condensed Matter 8 (1996) 339.

\bibitem{MineevSamokhin2008}
V.~P. Mineev, K.~V. Samokhin, Helical phases in superconductors, Physical
  Review B 78 (2008) 144503.

\bibitem{Agterberg2012}
D.~F. Agterberg, Magnetoelectric effects, helical phases, and {FFLO} phases, in
  Non-Centrosymmetric Superconductors: Introduction and Overview, edited by E.
  Bauer and M. Sigrist (Springer, Berlin) (2012) 155.

\bibitem{WakatsukiNagaosa2018}
R.~Wakatsuki, N.~Nagaosa, Nonreciprocal current in noncentrosymmetric {Rashba}
  superconductors, Physical Review Letters 121 (2018) 026601.

\bibitem{Hoshino2018}
S.~Hoshino, R.~Wakatsuki, K.~Hamamoto, N.~Nagaosa, Nonreciprocal charge
  transport in two-dimensional noncentrosymmetric superconductors, Physical
  Review B 98 (2018) 054510.

\bibitem{DaidoYanase2024}
A.~Daido, Y.~Yanase, Nonlinear paraconductivity in noncentrosymmetric
  superconductors, Physical Review Research 6 (2024) L022009.

\bibitem{Parafilo2022}
A.~V. Parafilo, V.~M. Kovalev, I.~G. Savenko, Photogalvanic transport in
  fluctuating {Ising} superconductors, Physical Review B 106 (2022) 144502.

\bibitem{Sonowal2023}
K.~Sonowal, A.~V. Parafilo, M.~V. Boev, V.~M. Kovalev, I.~G. Savenko,
  Second-harmonic generation in fluctuating {Ising} superconductors, 2D
  Materials 10 (2023) 045004.

\bibitem{BoevKovalev2024}
M.~V. Boev, V.~M. Kovalev, Photovoltaic {Hall} effect in a system of
  fluctuating {Cooper} pairs, JETP Letters 120 (2024) 494.

\bibitem{Parafilo2025}
A.~V. Parafilo, V.~M. Kovalev, I.~G. Savenko, Ratchet {Hall} effect in
  fluctuating superconductors, Physical Review B 112 (2025) 104501.

\bibitem{Levchenko2026}
A.~Levchenko, Effect of superconducting fluctuations on nonreciprocal dichroism
  and gyrotropy, Phys. Rev. B 114 (2026) 194506.

\bibitem{TdM2026a}
J.~T. de~Miranda, M.~Khodas, A.~Levchenko, Magnetochiral anisotropy in strained
  superconducting transition metal dichalcogenides, Phys. Rev. B 114 (2026)
  154503.

\bibitem{TdM2026b}
J.~T. de~Miranda, M.~Khodas, A.~Levchenko, Electrical magnetochiral anisotropy
  in {Rashba} superconductors, Low Temperature Physics 52~(9) (2026)
  1107--1112.

\bibitem{Levchenko2026TAPS}
A.~Levchenko, Magnetochiral anisotropy of thermally activated phase slips in
  noncentrosymmetric superconductors (2026).
\newblock \href {http://arxiv.org/abs/2609.08237} {\path{arXiv:2609.08237}}.

\bibitem{Rikken2001}
G.~L. J.~A. Rikken, J.~F{\"o}lling, P.~Wyder, Electrical magnetochiral
  anisotropy, Physical Review Letters 87 (2001) 236602.

\bibitem{TokuraNagaosa2018}
Y.~Tokura, N.~Nagaosa, Nonreciprocal responses from non-centrosymmetric quantum
  materials, Nature Communications 9 (2018) 3740.

\bibitem{IdeueIwasa2021}
T.~Ideue, Y.~Iwasa, Symmetry breaking and nonlinear electric transport in van
  der {Waals} nanostructures, Annual Review of Condensed Matter Physics 12
  (2021) 201.

\bibitem{Ando2020}
F.~Ando, Y.~Miyasaka, T.~Li, J.~Ishizuka, T.~Arakawa, Y.~Shiota, T.~Moriyama,
  Y.~Yanase, T.~Ono, Observation of superconducting diode effect, Nature 584
  (2020) 373.

\bibitem{DaidoIkedaYanase2022}
A.~Daido, Y.~Ikeda, Y.~Yanase, Intrinsic superconducting diode effect, Physical
  Review Letters 128 (2022) 037001.

\bibitem{Nadeem2023}
M.~Nadeem, M.~S. Fuhrer, X.~Wang, The superconducting diode effect, Nature
  Reviews Physics 5 (2023) 558.

\bibitem{Shaffer2025}
D.~Shaffer, A.~Levchenko, Theories of superconducting diode effects (2025).
\newblock \href {http://arxiv.org/abs/2510.25864} {\path{arXiv:2510.25864}}.

\bibitem{AslamazovLarkin1968}
L.~G. Aslamazov, A.~I. Larkin, The influence of fluctuation pairing of
  electrons on the conductivity of normal metal, Physics Letters A 26 (1968)
  238.

\bibitem{Schmid1966}
A.~Schmid, A time dependent {Ginzburg-Landau} equation and its application to
  the problem of resistivity in the mixed state, Physik der kondensierten
  Materie 5 (1966) 302.

\bibitem{Maki1968}
K.~Maki, The critical fluctuation of the order parameter in type-{II}
  superconductors, Progress of Theoretical Physics 39 (1968) 897.

\bibitem{Thompson1970}
R.~S. Thompson, Microwave, flux flow, and fluctuation resistance of dirty
  type-{II} superconductors, Physical Review B 1 (1970) 327.

\bibitem{Dong2025}
Z.-H. Dong, H.~Yang, Y.~Zhang, Enhanced nonlinear {Hall} effect by {Cooper}
  pairs near the superconducting phase transition, Physical Review B 111 (2025)
  155120.

\bibitem{LarkinVarlamov2005}
A.~Larkin, A.~Varlamov, Theory of Fluctuations in Superconductors, Oxford
  University Press, Oxford, 2005.

\bibitem{Liu2026}
T.~Liu, J.~T. de~Miranda, D.~Shaffer, A.~Levchenko, Kinetic {Lifshitz}
  invariants and dynamics of nonreciprocal fluctuations in superconductors
  (2026).
\newblock \href {http://arxiv.org/abs/2608.05306} {\path{arXiv:2608.05306}}.

\bibitem{ChaikinLubensky1995}
P.~M. Chaikin, T.~C. Lubensky, Principles of Condensed Matter Physics,
  Cambridge University Press, Cambridge, 1995.

\bibitem{GE1968}
L.~P. Gor'kov, G.~M. Eliashberg, Generalization of the {Ginzburg-Landau}
  equations for non-stationary problems in the case of alloys with paramagnetic
  impurities, Zh. Eksp. Teor. Fiz. 54 (1968) 612, [Sov. Phys. JETP \textbf{27},
  328 (1968)].

\bibitem{LevchenkoKamenev2007}
A.~Levchenko, A.~Kamenev, Keldysh {Ginzburg-Landau} action of fluctuating
  superconductors, Physical Review B 76 (2007) 094518.

\bibitem{Kamenev2011}
A.~Kamenev, Field Theory of Non-Equilibrium Systems, Cambridge University
  Press, Cambridge, 2011.

\bibitem{Zachmann2013}
M.~Zachmann, M.~D. Croitoru, A.~Vagov, V.~M. Axt, T.~Papenkort, T.~Kuhn,
  Ultrafast terahertz-field-induced dynamics of superconducting bulk and
  quasi-{1D} samples, New Journal of Physics 15 (2013) 055016.

\bibitem{Croitoru2023}
M.~D. Croitoru, B.~Lounis, A.~I. Buzdin, Helicity-controlled switching of
  superconducting states by radiation pulse, Applied Physics Letters 123 (2023)
  122601.

\end{thebibliography}

\end{document}